\newif\iffigs\figstrue
% Uncomment the next line if you do not want the figures
%\figsfalse

%
% the following is to use blackboard bold fonts --
\let\useblackboard=\iftrue
%
% activate this if you don't have them.
%\let\useblackboard=\iffalse
%
% You might also need to remove this line.
\newfam\black

\input harvmac.tex
\noblackbox

\iffigs
\input epsf
\else
  \message{No figures will be included.  See TeX file for more
information.}
\fi

\font\ninett=cmtt9

\ifx\answ\bigans\def\tcbreak#1{}\else\def\tcbreak#1{\cr&{#1}}\fi
\useblackboard
\message{If you do not have msbm (blackboard bold) fonts,}
\message{change the option at the top of the tex file.}
\font\blackboard=msbm10 %scaled \magstep1
\font\blackboards=msbm7
\font\blackboardss=msbm5
%\newfam\black
\textfont\black=\blackboard
\scriptfont\black=\blackboards
\scriptscriptfont\black=\blackboardss
\def\Bbb#1{{\fam\black\relax#1}}
\else
\def\Bbb#1{{\bf #1}}
\fi
\def\yboxit#1#2{\vbox{\hrule height #1 \hbox{\vrule width #1
\vbox{#2}\vrule width #1 }\hrule height #1 }}
\def\fillbox#1{\hbox to #1{\vbox to #1{\vfil}\hfil}}
\def\ybox{{\lower 1.3pt \yboxit{0.4pt}{\fillbox{8pt}}\hskip-0.2pt}}
\def\np#1#2#3{Nucl. Phys. {\bf B#1} (#2) #3}
\def\pl#1#2#3{Phys. Lett. {\bf #1B} (#2) #3}

\def\cmp#1#2#3{Comm. Math. Phys. {\bf #1} (#2) #3}

\def\comments#1{}

\def\half{{1\over 2}}

\def\Re{{\rm Re\hskip0.1em}}
\def\Im{{\rm Im\hskip0.1em}}

\def\CM{{\cal M}}

\def\CL{{\cal L}}

\def\a{\alpha}

\def\II{\relax{I\kern-.07em I}}

\def\IZ{\relax\ifmmode\mathchoice
{\hbox{\cmss Z\kern-.4em Z}}{\hbox{\cmss Z\kern-.4em Z}}
{\lower.9pt\hbox{\cmsss Z\kern-.4em Z}}
{\lower1.2pt\hbox{\cmsss Z\kern-.4em Z}}\else{\cmss Z\kern-.4em
Z}\fi}
\def\IB{\relax{\rm I\kern-.18em B}}
\def\IC{\bf C}
\def\ID{\relax{\rm I\kern-.18em D}}
\def\IE{\relax{\rm I\kern-.18em E}}
\def\IF{\relax{\rm I\kern-.18em F}}
\def\IG{\relax\hbox{$\inbar\kern-.3em{\rm G}$}}
\def\IGa{\relax\hbox{${\rm I}\kern-.18em\Gamma$}}
\def\IH{\relax{\rm I\kern-.18em H}}
\def\II{\relax{\rm I\kern-.18em I}}
\def\IK{\relax{\rm I\kern-.18em K}}
\def\IP{\relax{\rm I\kern-.18em P}}
%\def\IX{\relax{\rm X\kern-.01em X}}
%this doesn't work

\useblackboard
\def\IZ{\relax\Bbb{Z}}
\fi

\font\cmss=cmss10 \font\cmsss=cmss10 at 7pt
\def\IR{\relax{\rm I\kern-.18em R}}

\def\Re{{\rm Re\ }}
\def\Im{{\rm Im\ }}

\def\BZ{\IZ}

\def\bR{{\bf R}}
\def\bS{{\bf S}}
\def\bT{{\bf T}}
\def\tilde{\widetilde}

%%%%%%%%%%%%%%%%%%%%%%%%%%%%%%%%%%%%%%%%%%%%%%%%%%%%%%%%%%%%%%%%%%%%%%%%%%%%
%                    Definitions from LaTeX                                %
%%%%%%%%%%%%%%%%%%%%%%%%%%%%%%%%%%%%%%%%%%%%%%%%%%%%%%%%%%%%%%%%%%%%%%%%%%%%

                   % N=? SUSY
\def\lbr{{\lbrack}}                             % [
\def\rbr{{\rbrack}}                             % ]

\def\wdg{{\wedge}}                              % wedge product

                              % Wilson lines

\def\inv#1{{1\over{#1}}}                              % inverse
\def\ol#1{{{\cal O}({#1})}}                           % O(x)

               % Real numbers
               % Complex numbers

%%% \def\MR#1{{{\bf R}^{#1}}}               % Real numbers
%%% \def\MC#1{{{\bf C}^{#1}}}               % Complex numbers
               % Real numbers
               % Complex numbers
\def\MS#1{{{\bf S}^{#1}}}               % Circle, sphere,...
               % disk, ball,...
\def\MT#1{{{\bf T}^{#1}}}               % Torus
\def\CP#1{{{\bf CP}^{#1}}}              % CP
               % Ruled surface F_n

\def\tB{{\tilde B}}                     % self-dual B
                         % sign
\def\hepth#1{{\tt hep-th/{#1}}}

%%%%%%%%%%%%%%%%%%%%%%%%%%%%%%%%%%%%%%%%%%%%%%%%%%%%%%%%%%%%%%%%%%%%%%%%%%%%
%                    Greek                                                 %
%%%%%%%%%%%%%%%%%%%%%%%%%%%%%%%%%%%%%%%%%%%%%%%%%%%%%%%%%%%%%%%%%%%%%%%%%%%%

\def\g{{\gamma}}

%%%%%%%%%%%%%%%%%%%%%%%%%%%%%%%%%%%%%%%%%%%%%%%%%%%%%%%%%%%%%%%%%%%%%%%%%%%%
%                    TITLE PAGE                                            %
%%%%%%%%%%%%%%%%%%%%%%%%%%%%%%%%%%%%%%%%%%%%%%%%%%%%%%%%%%%%%%%%%%%%%%%%%%%%

%
\Title{ \vbox{\baselineskip12pt\hbox{hep-th/9610251,
       PUPT-1646, IASSNS-HEP-96/98, RU-96-86}}}
{\vbox{
\centerline{Branes, Calabi--Yau Spaces, and}\vskip6pt
\centerline{Toroidal Compactification of the}
\vskip6pt\centerline{$N{=}1$ Six-Dimensional $E_8$ Theory}
}}
\centerline{Ori J. Ganor,$^{1,a}$
David R. Morrison,$^{2,*,b}$
and Nathan Seiberg$^{3,c}$}
\bigskip
\centerline{$^1$Department of Physics, Jadwin Hall, Princeton
University, Princeton, NJ 08544}
\centerline{$^2$Schools of Mathematics and Natural
Sciences, Institute for Advanced Study,}\centerline{Princeton, NJ 08540}
\centerline{$^3$Department of Physics and Astronomy, Rutgers University,
Piscataway, NJ 08855-0849}
\bigskip
\bigskip
\bigskip

{\parindent=-5pt

\footnote{}{${}^*${On leave from: Department of Mathematics, Duke
University, Durham, NC 27708-0320}\par
${}^a${\ninett origa@puhep1.princeton.edu}\par
${}^b${\ninett drm@math.duke.edu}\par
${}^c${\ninett seiberg@physics.rutgers.edu}
}
}

\noindent
We consider compactifications of the $N=1$, $d=6$, $E_8$ theory on tori to
five, four, and three dimensions and learn about some properties of this
theory.  As a by-product we derive the $SL(2,\IZ)$ duality of the $N=2$,
$d=4$, $SU(2)$ theory with $N_f=4$.  Using this theory on a D-brane probe
we shed new light on the singularities of F-theory compactifications to
eight dimensions.  As another application we consider compactifications of
F-theory, M-theory and the IIA string on (singular) Calabi-Yau spaces where
our theory appears in spacetime.  Our viewpoint leads to a new perspective
on the nature of the singularities in the moduli space and their spacetime
interpretations. In particular, we have a universal understanding of how
the singularities in the classical moduli space of Calabi--Yau spaces are
modified by worldsheet instantons to singularities in the moduli space of
the corresponding conformal field theories.

\Date{October, 1996}

%%%%%%%%%%%%%%%%%%%%%%%%%%%%%%%%%%%%%%%%%%%%%%%%%%%%%%%%%%%%%%%%%%%%
%  B I B L I O G R A P H Y                                         %
%%%%%%%%%%%%%%%%%%%%%%%%%%%%%%%%%%%%%%%%%%%%%%%%%%%%%%%%%%%%%%%%%%%%

\lref\nclp{For a nice review see, S. Chaudhuri, C. Johnson, and J.
Polchinski, ``Notes on D-Branes,'' {\tt hep-th/9602052}.}

\lref\ardou{P.C. Argyres and M.R. Douglas,
   ``New Phenomena In $SU(3)$ Supersymmetric Gauge Theory,''
\np{448}{1995}{93--126}, \hepth{9505062}.}

\lref\apsw{P.C. Argyres, M.R. Plesser, N. Seiberg, and E. Witten,
  { ``New $N=2$ Superconformal Field Theories In Four Dimensions,''}
  \np{461}{1996}{71--84}, \hepth{9511154}.}

\lref\SWthreeD{N. Seiberg and E. Witten,
  { ``Gauge Dynamics And Compactification To Three Dimensions,''}
  \hepth{9607163}.}

\lref\gantc{O.J. Ganor,
  { ``Toroidal Compactification of Heterotic 6D
  Non-Critical Strings Down to Four Dimensions,''} \hepth{9608109}.}

\lref\swi{N. Seiberg and E. Witten,
  {``Electric-Magnetic Duality, Monopole Condensation,
  and Confinement in $N=2$ Supersymmetric Yang--Mills Theory,''}
  \np{426}{1994}{19--52}, \hepth{9407087}.}

\lref\swii{N. Seiberg and E. Witten,
  {``Monopoles, Duality and Chiral Symmetry Breaking in $N=2$
  Supersymmetric QCD,''} \np{431}{1994}{484--550}, \hepth{9408099}.}

\lref\GriHar{P. Griffiths and J. Harris,
  {\it Principles of Algebraic Geometry}, Wiley-Interscience, New York,
  1978.}

\lref\morsei{D.R. Morrison and N. Seiberg,
  {``Extremal Transitions and Five-Dimensional
   Supersymmetric Field Theories,''} \hepth{9609070}.}

\lref\dkv{M.R. Douglas, S. Katz, and C. Vafa,
  {``Small Instantons, del Pezzo Surfaces and Type I$'$ Theory,''}
  \hepth{9609071}.}

\lref\ganoha{O. Ganor and A. Hanany, ``Small E(8) Instantons and
Tensionless Noncritical Strings,'' \np{474}{1996}{122--140},
\hepth{9602120}.}

\lref\seiwit{N. Seiberg and E. Witten, ``Comments on String Dynamics in
Six Dimensions,'' \np{471}{1996}{121--134}, \hepth{9603003}.}

\lref\fived{N. Seiberg, ``Five Dimensional SUSY Field Theories,
Non-Trivial Fixed Points, and String Dynamics,'' {\tt hep-th/9608111}.}

\lref\mandf{E. Witten, ``Phase
Transitions in M-Theory and F-Theory,'' \np{471}{1996}{195--216}, {\tt
hep-th/9603150}.}

\lref\threedone{N. Seiberg, ``IR Dynamics on Branes and Space-Time
Geometry,'' {\tt hep-th/9606017}.}

\lref\nonren{N. Seiberg, ``Naturalness Versus Supersymmetric
Non-Renormalization Theorems,'' Phys. Lett. {\bf B318} (1993) 469--475,
{\tt hep-ph/9309335}.}

\lref\aps{P.C. Argyres, M.R. Plesser, and N. Seiberg,
``The Moduli Space of Vacua of $N=2$ SUSY QCD and Duality in $N=1$ SUSY
QCD,'' \np{471}{1996}{159--194}, {\tt hep-th/9603042}.}

\lref\vafaf{C. Vafa, ``Evidence for F-Theory,''
\np{469}{1996}{403--418}, \hepth{9602022}.}

\lref\MVI{D.R. Morrison and C. Vafa, ``Compactifications of F-Theory
on Calabi--Yau Threefolds -- I,'' \np{473}{1996}{74--92},
\hepth{9602114}.}

\lref\MVII{D.R. Morrison and C. Vafa, ``Compactifications of F-Theory on
Calabi--Yau Threefolds (II),'' \np{476}{1996}{437--469}, {\tt
hep-th/9603161}.}

\lref\bds{T. Banks, M.R. Douglas, and N. Seiberg, ``Probing $F$-Theory
With Branes,'' \pl{387}{1996}{278--281}, \hepth{9605199}.}

\lref\intse{K. Intriligator and N. Seiberg, ``Mirror Symmetry in Three
Dimensional Gauge Theories,'' \pl{387}{1996}{513--519},
\hepth{9607207}.}

\lref\minnem{J.A. Minahan and D. Nemeschansky, ``An $N=2$ Superconformal
Fixed Point with $E_6$ Global Symmetry,'' \hepth{9608047}.}

\lref\minnemtwo{J.A. Minahan and D. Nemeschansky, ``Superconformal Fixed
Points with $E_n$ Global Symmetry,'' \hepth{9610076}.}

\lref\lerwar{W. Lerche and N.P. Warner,
  { ``Exceptional SW Geometry from ALE Fibrations,''}
  \hepth{9608183}.}

\lref\seibergsix{N. Seiberg, ``Non-Trivial Fixed Points of The
Renormalization Group in Six Dimensions,'' \hepth{9609161}.}

\lref\wittensix{E. Witten, ``Physical Interpretation Of Certain Strong
Coupling Singularities,'' \hepth{9609159}.}

\lref\kodaira{K. Kodaira, ``On Compact Analytic Surfaces, II, III,''
Annals of Math. (2) {\bf 77} (1963) 563--626;
Annals of Math. (2) {\bf 78} (1963) 1--40.}

\lref\kmv{A. Klemm, P. Mayr, and C. Vafa, ``BPS States of Exceptional
Non-Critical Strings,'' {\tt hep-th/9607139}.}

\lref\kkv{S. Katz, A. Klemm, and C. Vafa, ``Geometric Engineering of
Quantum Field Theories,'' \hepth{9609239}.}

\lref\nekrasov{N. Nekrasov, ``Five Dimensional Gauge Theories and
Relativistic Integrable Systems,'' \hepth{9609219}.}

\lref\tate{J. Tate, ``Algorithm for
Determining the Type of a Singular Fiber in an Elliptic Pencil,'' in
{\it Modular Functions of One Variable IV}, Lecture Notes in
Math. vol. 476, Springer-Verlag, Berlin (1975).}

\lref\bikmsv{M. Bershadsky, K. Intriligator, S. Kachru, D. R. Morrison,
 V. Sadov, and C. Vafa, ``Geometric Singularities and Enhanced Gauge
Symmetries,''  Nucl. Phys. B, to appear,
 \hepth{9605200}.}

\lref\rKMP{S.~Katz, D.R. Morrison, and
M.R. Plesser, ``Enhanced Gauge Symmetry in Type II String Theory,''
\np{477}{1996}{105--140}, {\tt hep-th/9601108}.}

\lref\BKKM{P. Berglund, S. Katz,
A. Klemm, and P. Mayr, ``New Higgs Transitions Between Dual $N=2$ String
Models,'' {\tt hep-th/9605154}, and private communication.}

\lref\phases{E. Witten, ``Phases of $N{=}2$ Theories in Two
Dimensions,'' \np{403}{1993}{159--222}, \hepth{9301042}.}

\lref\small{P.S. Aspinwall, B.R. Greene, and D.R. Morrison, ``Measuring
Small Distances in $N{=}2$ Sigma Models,'' \np{420}{1994}{184--242},
\hepth{9311042}.}

\lref\HKTY{S. Hosono, A. Klemm, S. Theisen, and S.-T. Yau, ``Mirror
Symmetry, Mirror Map, and Applications to Calabi--Yau Hypersurfaces,''
\cmp{167}{1995}{301--350}, \hepth{9308122}.}

\lref\twoparamtwo{P. Candelas, A. Font, S. Katz, and D.R. Morrison,
``Mirror Symmetry for Two
Parameter Models (II),'' \np{429}{1994}{626--674}, \hepth{9403187}.}

\lref\nikulin{V.V. Nikulin,
``Integer Symmetric Bilinear Forms and Some of Their Geometric
Applications (Russian),'' Izv. Akad. Nauk SSSR Ser. Mat. {\bf 43} (1979)
111--177; English translation: Math. USSR Izvestija {\bf 14} (1980)
103--167.}

\lref\mirmor{R. Miranda and D.R. Morrison, ``The Number of Embeddings of
Integral Quadratic Forms, I, II,'' Proc. Japan Acad. Ser. A Math. Sci.
{\bf 61} (1985) 317--320; {\bf 62} (1986) 29--32.}

\lref\aspgr{P.S. Aspinwall and M. Gross, ``The $SO(32)$ Heterotic String
on a K3 Surface,'' \pl{387}{1996}{735--742}, \hepth{9605131}.}

\lref\burnsrap{D. Burns, Jr. and M. Rapaport, ``On the Torelli Problem
for
K\"ahlerian K3 Surfaces,'' Ann. Sci. \`Ecole Norm. Sup.
(4) {\bf 8} (1975) 235--274.}

\lref\wilson{P.M.H. Wilson, ``The K\"ahler Cone on Calabi--Yau
Threefolds,'' Invent. Math. {\bf 107} (1992) 561--583; Erratum,
ibid. {\bf 114} (1993) 231--233.}

\lref\kachvaf{S. Kachru and C. Vafa, ``Exact Results for $N{=}2$
Compactifications of Heterotic Strings,'' \np{450}{1995}{69--89},
\hepth{9505105}.}

\lref\FHSV{S. Ferrara, J.A. Harvey, A. Strominger, and C. Vafa,
``Second-Quantized Mirror Symmetry,'' \pl{361}{1995}{59--68},
\hepth{9505162}.}

\lref\KKLMV{S. Kachru, A. Klemm, W. Lerche, P. Mayr, and C. Vafa,
``Nonperturbative Results on the Point Particle Limit of $N{=}2$
Heterotic String Compactifications,'' \np{459}{1996}{537--558},
\hepth{9508155}.}

\lref\GHL{C. G\'omez, R. Hern\'andez, and E. L\'opez, ``$S$-Duality and
the Calabi--Yau Interpretation of the $N{=}4$ to $N{=}2$ Flow,''
\hepth{9512017}.}

\lref\asplouis{P.S. Aspinwall and J. Louis, ``On the Ubiquity of K3
Fibrations in String Duality,'' \pl{369}{1996}{233--242},
\hepth{9510234}.}

\lref\aspgauge{P.S. Aspinwall, ``Enhanced Gauge Symmetries and
Calabi--Yau Threefolds,'' \pl{371}{1996}{231--237}, \hepth{9511171}.}

\lref\ccdf{A.C. Cadavid, A. Ceresole, R. D'Auria, and S. Ferrara,
``Eleven-Dimensional Supergravity Compactified on a Calabi--Yau
Threefold,''
Phys. Lett. {\bf B357} (1995) 76--80, {\tt hep-th/9506144}.}

\lref\fkm{S. Ferrara,
R.R. Khuri, and R. Minasian, ``M-Theory on a Calabi--Yau Manifold,''
Phys. Lett. {\bf B375} (1996) 81--88,  {\tt hep-th/9602102}.}

\lref\fms{S. Ferrara,
R. Minasian, and A. Sagnotti, ``Low Energy Analysis of $M$ and $F$
Theories on Calabi--Yau Threefolds,'' \np{474}{1996}{323--342}, {\tt
hep-th/9604097}.}

\lref\intseirev{K. Intriligator and N. Seiberg,
``Lectures on Supersymmetric Gauge Theories and Electric-Magnetic
Duality,'' \hepth{9509066}.}

\lref\wittenII{E. Witten, ``Some Comments on String Dynamics,''
Contributed to STRINGS 95: Future Perspectives in String Theory, Los
Angeles, CA, 13-18 Mar 1995, \hepth{9507121}.}

\lref\bound{E. Witten, ``Bound States Of Strings And $p$-Branes,''
\np{460}{1996}{335--350}, \hepth{9510135}.}

\lref\BSV{M. Bershadsky, V. Sadov, and C. Vafa, ``D-Strings on
D-Manifolds,'' \np{463}{1996}{398--414}, \hepth{9510225}.}

\lref\Sen{A. Sen, ``$F$-Theory and Orientifolds,''
\np{475}{1996}{562--578}, \hepth{9605150}.}

\lref\rDM{K. Dasgupta and S. Mukhi, ``F-Theory at Constant Coupling,''
\pl{385}{1996}{125--131}, \hepth{9606044}.}

%%%%%%%%%%%%%%%%%%%%%%%%%%%%%%%%%%%%%%%%%%%%%%%%%%%%%%%%%%%%%%%%%%%%
%  Introduction                                                    %
%%%%%%%%%%%%%%%%%%%%%%%%%%%%%%%%%%%%%%%%%%%%%%%%%%%%%%%%%%%%%%%%%%%%
\newsec{Introduction}

Recent advances in local quantum field theories have uncovered a very
rich structure.  In particular, non-trivial fixed points in various
dimensions have been found both in theories with four supersymmetries
(for a review and earlier references see
\intseirev)
and with eight supersymmetries \refs{\ardou, \apsw, \fived, \morsei,
\minnem, \lerwar, \minnemtwo}.
Also, it turns out to be interesting to study
compactifications of a field theory as a function of the parameters of
the compactification, thus interpolating between field theories in
various dimensions.

One of the goals of this paper is to study the compactification on a
torus of the simplest non-trivial fixed point with $N=1$ supersymmetry
in six dimensions (for a recent discussion of other fixed points in
six dimensions see \refs{\seibergsix, \wittensix}).  Its global
symmetry is $E_8$ and it was first found in the study of
small $E_8$ instantons in string theory \refs{\ganoha, \seiwit}.  The
compactified theory depends on various parameters: the moduli of the
torus and twists in the boundary conditions which break $E_8$ to its
subgroups.

Our analysis proceeds in parallel from three different points of view:

\item{1.}  The $d=6$ $E_8$ theory is still mysterious.  Although it
looks like a local quantum field theory, it does not have a Lagrangian
description.  We expect that by studying the properties of this theory a
simple presentation of the theory which makes its behavior manifest will
emerge.  Hopefully, such a presentation will also be useful in other
field theories.

\item{2.}  The $d=6$ $E_8$ theory is the low energy description of
five-branes in M-theory near the ``end-of-the-world'' brane.  When this
eleven-dimensional theory and the five-brane are compactified on a torus,
the low energy theory on the brane is the theory mentioned above.  Using
dualities, this is the theory on lower dimension probes in
compactifications of the type I$'$ on $\bS^1/\IZ_2$ \fived, F-theory
\vafaf\ on K3 \bds\ and M-theory on K3 \threedone.  The
compactified $d=6$ $E_8$ theory thus tells us about the behavior of
these compactifications.

\item{3.}  The $d=6$ $E_8$ theory is the low energy theory of the
heterotic theory compactified on K3 near the limit as an $E_8$
instanton in the compactification shrinks to zero size.  Another
description of this theory is the compactification of F-theory on a
singular Calabi--Yau space \refs{\MVI, \seiwit, \MVII, \mandf}.  After
compactification on another circle, this becomes the low energy theory
of M-theory compactified on the same Calabi--Yau space.
Varying parameters to break $E_8$ to its subgroups will alter the
singularity type of the Calabi--Yau space.  After
further compactification on a circle we find a compactification of the
type IIA theory on the same singular Calabi--Yau space.  Therefore, an
understanding of this field theory can be achieved by relating it
to the singularities of
the Calabi--Yau compactifications and the way they are corrected by
worldsheet instanton effects.  Note that this field theoretic
understanding of these compactifications makes it obvious that the
nature of the singularities in the conformal field theory moduli space
is independent of the details of the underlying Calabi--Yau space;
they depend only on the singularities of that space.

Although these three applications are distinct, we will often find it
easier to make an argument based on one point of view than on the
others.  Then we can translate the conclusion to learn about the other
applications.

The six-dimensional theory has no free parameters.  The moduli space of
vacua has two branches.  The Coulomb branch is $\bR^+$.  The real scalar
field which parameterizes it is in a tensor multiplet.  There is also a
Higgs branch isomorphic to the moduli space of $E_8$ instantons.

In lower dimensions new parameters, which are associated with the
compactification appear.  In five dimensions one real parameter is
$R_6$---the radius of the circle.  We can also couple the $E_8$
currents of the global symmetry to background gauge fields.  This leads
to eight more real parameters (the Wilson lines around the circle).  It
is convenient to think of them as background superfields \nonren.  In
this case, all of them are in vector multiplets \aps\ of the
five-dimensional theory.  Specifically, $1 \over R_6$ is the scalar of a
background vector superfield.

In four dimensions we get more parameters. First, the eight scalars
in the $E_8$ background gauge fields become complex.  Second, the
compactified $\bT ^2$ leads to two other vector superfields.  Three of
the four real scalars in these superfields are the two sides of the
torus $R_5$ and $R_6$, and the angle between them $\varphi$.  The
fourth one, which is needed for supersymmetry, is a background $\tilde
B_{56}$ field.  $\tilde B_{\mu\nu}$ is a background two-form whose
three-form field strength is self-dual.  It is a component of the
gravity multiplet in six dimensions.  More explicitly,
\eqn\vvi{\eqalign{
v_5 &= {1\over 2\sqrt{2}\pi R_5\sin\varphi} e^{-i\varphi/2+i\tB/2} \cr
v_6 &= {1\over 2\sqrt{2}\pi R_6\sin\varphi} e^{i\varphi/2+i\tB/2} \cr}}
are the scalar components of background vector superfields.

Since all these parameters are components of vector superfields, the way
they affect the solutions is very restricted.  First, the Higgs branch
is independent of their values.  The only dependence follows {}from
symmetry breaking---when the eight parameters break $E_8$ to a subgroup
$G$, the Higgs branch is the moduli space of $G$ instantons (when $G$ is
a product of several simple factors, there are several distinct Higgs
branches).  Second, in five dimensions the coupling of these vectors is
very limited and essentially determined.  In four dimensions there is
more freedom than in five dimensions.  But the constraints of
supersymmetry are still useful---they imply that the elliptic curve
describing the Coulomb branch \swi\ varies holomorphically in these
parameters.

In section 2 we consider the compactification to five dimensions.  We
will see how the theories discussed in \refs{\fived, \morsei, \dkv} are
obtained.  Also, we will show that there is always one more singularity
in the moduli space and we will discuss its interpretation.  In terms of
the six dimensional theory it arises {}from a string with winding number
$L_6=1$ and momentum $P_6=1$ which becomes massless at that point.  We
will discuss this singularity both {}from the point of view of
the theory on the brane and {}from the point of view of M-theory
compactification.

In section 3 we initiate a study of compactification to four dimensions
following \gantc.  We first restrict ourselves to the subspace of the
parameter space preserving the $E_8$ symmetry and study its properties.
We then break the $E_8$ symmetry with Wilson lines and discuss more
general situations.  For special values of the Wilson lines we recover
the $N=2$, $d=4$, $SU(2)$ theory with $N_f=4$.  Our six-dimensional
viewpoint leads to a derivation of its $SL(2,\IZ)$ duality \swii\ as a
manifestation of the $SL(2,\IZ)$ which acts on the torus we compactify
on.

In section 4 we study the singularities of the four-dimensional theory
in more detail.  The possible singularities were classified by Kodaira
\kodaira.  We identify each of them with a four-dimensional field
theory.  Then, we consider the compactification of the various
five-dimensional field theories on a circle and map them to
singularities in four dimensions.

In section 5 we use the results of section 4 in the context of the
theory on a three-brane in F-theory compactification to eight dimensions
and in the context of type IIA compactifications to four dimensions.  In
particular, we shed new light on the behavior of the moduli space of the
conformal field theories on the string worldsheet as corrected by
worldsheet instantons near singularities of the geometry. Some of these
singularities are understood as a result of non-trivial dynamics in a
four-dimensional field theory.  The short distance degrees of freedom of
this field theory are visible in five dimensions.

In section 6 we make some comments about the compactification to three
dimensions. Finally, in the appendices we give some more technical
details.

\newsec{From six to five dimensions}

We start with the six-dimensional $N=1$ supersymmetric
field theory associated with small $E_8$ instantons.  It has $E_8$
global symmetry.
Being at a fixed point of the renormalization group it is exactly scale
invariant.  The Higgs branch of the theory is isomorphic to the moduli
space of $E_8$ instantons.  It is a hyper-K\"ahler manifold with $E_7$
symmetry and $E_8$ action.  The massless hypermultiplets transform as
$\half ({\bf 56}) +{\bf 1}$ under $E_7$ where the $\half$ stands for
half hypermultiplets.

The Coulomb branch of the theory is $\bf R^+$.  The low energy degrees
of freedom along this Coulomb branch are in a tensor multiplet which
includes a two-form with self-dual field strength, a scalar $\Phi$ and a
fermion.  The kinetic term for $\Phi$ determines the metric on the
Coulomb branch
\eqn\sixdla{\CL_6 = \inv{32\pi} (\partial \Phi)^2.}

Note that $\Phi$ has dimension two and this effective Lagrangian (like
the full theory) is scale invariant.  More precisely, scale
invariance is spontaneously broken along the Coulomb branch.  The
expectation value of $\Phi$ determines the tension of BPS strings
\eqn\sixdstring{T= \sqrt{2}Z = \Phi.}
Arguments based on string theory show that such BPS strings exist and
they carry a chiral $E_8$ current algebra.

Since parameters are introduced in these theories by coupling them to
background vector superfields, and since those do not have scalars,
these quantum field theories do not have relevant operators which
preserve the super Poincar\'e symmetry.

Let us compactify this theory on $\MS{1}$ of radius $R_6$ to five
dimensions. For large $\Phi\gg {1 \over R_6^2}$ the massive modes
decouple {}from the light modes.  Therefore, in this regime the
five-dimensional theory is obtained by dimensional reduction of \sixdla
\eqn\fivdla{\CL_5 = \inv{16} R_6 (\partial \Phi)^2 ~~{\rm for} ~~
\Phi \gg {1 \over R_6^2}.}
The natural parameter on the Coulomb branch in five dimensions has
dimension one.  It is
\eqn\defphi{\phi= \sqrt{2}\pi R_6\Phi}
and \fivdla\ becomes
\eqn\fivdlan{\CL_5 = {1\over 32\pi^2 R_6} (\partial \phi)^2 ~~{\rm
for} ~~ \phi \gg {1 \over R_6}.}
Similarly we can reduce the rest of the Lagrangian.  The two-form
becomes a gauge field with a $\phi$-independent coupling constant
\eqn\gaugelarp{t(\phi)= {16 \pi^2 \sqrt{2} \over g_{eff}^2(\phi) } =
{\sqrt{2} \over R_6} ~~{\rm for} {}~~ \phi \gg {1 \over R_6}.}
Our normalization is such that ${8\pi \over g_{eff}^2(\phi) } = {\partial
\phi_D \over \partial \phi}$ and $\phi_D \approx \Phi/\sqrt{2}$,
$\phi\approx \sqrt{2}\pi R_6 \Phi$ for $\phi \gg {1 \over R_6}$.  This
definition of $g_{eff}$ differs by a multiplicative factor {}from the
one used in \fived; $t(\phi)$ is as in \morsei.

At small $\phi$ the five-dimensional theory has been analyzed \fived\
with the result
\eqn\gaugesmp{t(\phi)=2 \phi ~~{\rm for} ~~
0\le \phi \ll {1 \over R_6}.}
Note that this behavior is consistent with the scale invariance of the
five-dimensional theory at the origin at long distance.
The two asymptotics \gaugelarp\ and \gaugesmp\ must be sewed in a way
consistent with the restrictions {}from supersymmetry.  The slope
${\partial t(\phi)\over \partial \phi }$
can have discontinuities which are
a positive integer multiple of $-2$.  Therefore, the only solution is
\eqn\gefffs{t(\phi) = \cases{
2\phi& ~~ for~~ $0<\phi< {1 \over \sqrt{2} R_6}$ \cr
{\sqrt{2} \over R_6}& ~~for~~  ${1 \over \sqrt{2}  R_6} <\phi$ \cr}}
That transition is exactly the one associated with a single
hypermultiplet with electric charge one whose mass is
$\sqrt{2}|\phi - {1 \over \sqrt{2}R_6}|$.
The shift by ${1 \over R_6}$ in the mass indicates that
the central charge for particles is
\eqn\centralch{Z=n_e \phi - {P_6 \over \sqrt{2} R_6}}
where $n_e$ is the electric charge and $P_6$ is a charge of a global
symmetry.  The $R_6$ dependence of $Z$ suggests that it can be
interpreted as momentum in the compact direction.  The particle which
leads to the transition has $n_e=P_6=1$ and for consistency we have to
assume that there are no other particles which become massless on the
Coulomb branch away {}from $\phi=0$.

Better understanding of the theory in six dimensions that we started
with will have to explain why it has exactly one such particle with
these quantum numbers.  In the meantime, we note that since the
electric charge $n_e$ arises as the winding number of the
six-dimensional string around the circle, our state has $L_6=P_6=1$.
States with $L_6=P_6=n$ must be interpreted as multiple states rather
than as other elementary states in five dimensions.  Since our state
behaves as an ordinary particle in five dimensions, it can have
arbitrary five-momentum.  Hence we can summarize the restriction on
the quantum numbers by
\eqn\sumstates{\vec L \cdot \vec P =1.}
This is the equation which will have to be explained by better
understanding of the six-dimensional theory.

Using \gefffs\ it is easy to find
\eqn\phid{\phi_D(\phi) = \cases{
{\phi^2 \over 2\sqrt{2}\pi } & ~~for~~ $0<\phi< {1 \over \sqrt{2} R_6}$
\cr
{\phi \over 2\pi R_6} - {1 \over 4\sqrt{2}\pi R_6^2}& ~~for~~
${1 \over \sqrt{2} R_6} <\phi$
\cr}}
where the integration constant was set such that $\phi_D$ is continuous.
We see that for ${1 \over \sqrt{2} R_6} <\phi$ strings have the tension
$(\Phi - {1 \over 4 \pi R_6^2})$---it is as in the six-dimensional
theory up to a finite shift which vanishes as $R_6 \rightarrow \infty$.
For $0<\phi< {1 \over \sqrt{2}R_6}$ the tension of strings is
$\phi^2/2\pi$ and they become tensionless at $\phi=0$.

Equation \gefffs\ has a natural interpretation as the coupling constant
in the effective theory on a D4-brane \nclp\ in string theory.  It
describes the interaction of a D4-brane probe with a background
orientifold at $\phi=0$ where the coupling diverges \fived.  Since the
effective coupling for $\phi > {1 \over \sqrt{2}R_6}$ is constant, we
can add arbitrarily far away, around $\phi = \phi_0 \gg {1 \over
\sqrt{2}R_6}$, eight more background D8-branes and an orientifold to
have a compactification of the type IIA theory on an orientifold
$\MS{1}/\BZ_2$.  This theory is equivalent to M-theory compactified on
$\MS{1} \times (\MS{1}/\BZ_2)$ where the two radii are $R_6$ and $
\phi_0/\pi $.  In the limit of large $\phi_0$ our theory focuses on the
vicinity of one end-of-the-world brane in M-theory compactified on
$\MS{1}$ of radius $R_6$.  The effective theory on a five-brane of
M-theory wrapped on the $\MS{1}$ is the theory described above.

This analysis clarifies the behavior of the corresponding tensionless
string in F-theory.  Suppose we have a six-dimensional F-theory model
with an almost-tensionless string which in the limit of zero tension is
described by the $E_8$ theory.  Concretely, this means that the base of
the elliptic fibration contains a two-sphere $\Sigma$ of
self-intersection $-1$ whose area goes to zero in the limit.  When this
F-theory model is compactified on a circle, the resulting
five-dimensional theory is dual to M-theory on the total space of the
elliptic fibration.  The almost-tensionless string in six dimensions
gives rise to both a particle and a string in five dimensions, depending
on whether we wrap it on the circle or not. {}From the M-theory
perspective, the string comes {}from wrapping the five-brane around the
four-manifold $S$ which lies over $\Sigma$, and the particle comes
{}from wrapping the two-brane around a two-manifold (isomorphic to
$\Sigma$) within the four-manifold.

As shown in \MVII\ and further discussed in \kmv, in the M-theory model
with any nonzero value of $1/R_6$, as the area of $\Sigma$ goes to zero,
the volume of $S$ remains positive.  The massless particle at the
zero-area limit is the signal of a flop transition, and indeed the total
space can be flopped.  Continuing further in the moduli space, the
volume of $S$ can then be shrunk to zero.  In other words, {\it the
particle becomes massless at a different parameter value {}from where
the string becomes tensionless.} This second transition is described by
the five-dimensional $E_8$ theory of \fived; since it is an interacting
theory, attempts to describe it in terms of free fields are likely to
lead to confusing results such as an infinite number of light states.
It is crucial to stress that there is nothing pathological about the low
energy theory.

By adding Wilson lines with values in $E_8$, the $E_8$ symmetry can be
broken.  Then, various different critical theories can be found \fived.
At $\phi\not=0$ the only special theories are those of $n$ massless
hypermultiplets.  We denote them by $A_{n-1}$.  Their symmetry is
$SU(n)$.  At $\phi=0$ we can have an $SU(2)$ theory with $n$
hypermultiplets which we denote by $D_n$.  Its symmetry is $SO(2n)$.
For $n=0$ there is a possibility of turning on a discrete $\theta$
parameter \dkv\ and therefore there are two theories $D_0$ and $\tilde
D_0$ differing only in their massive spectra \morsei.  If the coupling
constant diverges at $\phi=0$ we also find a series of $E_n$ theories
with $n=0,...,8$ with symmetry $E_n$: $E_8$, $E_7$, $E_6$,
$E_5=Spin(10)$, $E_4=SU(5)$, $E_3=SU(3)\times SU(2)$, $E_2=SU(2)\times
U(1)$ and $E_1=SU(2)$ ($E_0$ has no symmetry).  The final possibility is
the theory $\tilde E_1$ whose symmetry is $U(1)$.

As we said above, these theories are obtained in string
compactifications to nine dimensions.  The $A_n$ theories are obtained
on a D4-brane probe near $n+1$ coalescing D8-branes.  The $D_n$ theories
are obtained near an orientifold with $n$ D8-branes.  The $E_n$ theories
are obtained when the string coupling diverges at the orientifold.

The other application of these theories is the low energy limit of
M-theory compactified on a Calabi--Yau space $X$ to five dimensions.
They correspond to the various singularities of Calabi--Yau spaces
\refs{\mandf, \morsei, \dkv}.
The $A_n$ theories are obtained {}from $n+1$ rational curves in a single
homology class shrinking to a point.  The $D_n$ theories correspond to a
rational ruled surface $S$ with $n$ singular fibers, shrinking to a
rational curve.  In the generic situation, the components of the
singular fibers will all lie in a common homology class $\gamma$, and
the fiber itself will be in class $2\gamma$.  Finally, the $E_n$ cases
correspond to a del Pezzo surface $S$ shrinking to a point.  In the
generic situation, the image of the restriction map $H^{1,1}(X)\to
H^{1,1}(S)$ will be a one-dimensional space, spanned by $c_1(S)$.

%%%%%%%%%%%%%%%%%%%%%%%%%%%%%%%%%%%%%%%%%%%%%%%%%%%%%%%%%%%%%%%%%%%%
%  4D                                                              %
%%%%%%%%%%%%%%%%%%%%%%%%%%%%%%%%%%%%%%%%%%%%%%%%%%%%%%%%%%%%%%%%%%%%

\newsec{First look at compactification to four dimensions}

\subsec{Preliminaries}

In this section we begin an analysis of the compactification of the
$E_8$ theory to four dimensions on a two torus.  The generic low-energy
fields comprise a $U(1)$ vector multiplet of $N=2$ in four dimensions.
The Lagrangian can be constructed \gantc\ {}from an elliptic curve as in
\swi.

The $\MT{2}$ torus has sides $R_5, R_6$ and angle $\varphi$.
We define, as in \vvi,
\eqn\vv{\eqalign{
v_5 &= {1\over 2\sqrt{2}\pi R_5\sin\varphi} e^{-i\varphi/2+i\tB/2} \cr
v_6 &= {1\over 2\sqrt{2}\pi R_6\sin\varphi}
e^{i\varphi/2+i\tB/2}. \cr}}
Its complex structure is
\eqn\sig{\eqalign{
\sigma &= {R_5\over R_6}e^{i\varphi}= {v_6 \over v_5} \cr
q &= e^{2\pi i\sigma}. \cr}}

We first consider the region far along the flat directions.  In that
region we can neglect all effects associated with massive modes and
simply dimensionally reduce the six dimensional free tensor
multiplet.  The bosonic fields of the tensor multiplet are $\Phi$ and
$B_{\mu\nu}^{(-)}$.  There is no Lagrangian for the anti-self-dual
two-form but we can start with
\eqn\lagr{L_{6D} = \inv{32\pi} (d\Phi)^2 + \inv{32\pi}(dB)^2 }
and set the self-dual part of $B$ to zero.
The coupling constant for $B$ is determined by
anti-self-duality. In 4D we find the fields
\eqn\fields{\phi_1 = 2\pi R_6 \Phi,\qquad
A_\mu = B_{\mu 6},\qquad
\phi_2 = \inv{2\pi R_5} B_{56}.}
In terms of the field-strength $F_{\mu\nu} =
\partial_{\lbr\mu}A_{\nu\rbr}$ we have
\eqn\intermsoffs{\partial_{\lbr\mu}B_{\nu\rbr5}
={R_5 \sin\varphi\over R_6} \widetilde{F}_{\mu\nu}
+{R_5 \cos\varphi\over R_6} F_{\mu\nu}
= \Im\left\lbrack\sigma (\widetilde{F}_{\mu\nu}
                   + i F_{\mu\nu})\right\rbrack .}
The 4D Lagrangian is thus
\eqn\newlag{L_{4D} =
 {\Im\sigma\over 32\pi} (d\phi_1)^2
+ {\Im\sigma\over 32\pi \sin^2\varphi} (d\phi_2)^2
+ {\Im\sigma\over 32\pi} F^2
+ {\Re\sigma\over 32\pi} F\widetilde{F}.}
We define
\eqn\aad{\eqalign{
a =& \eta \inv{\sqrt{2}}\left(\phi_1 + {i\over
\sin\varphi}\phi_2\right),\cr
a_D =& \sigma a,\cr }}
where $\eta$ is an arbitrary phase which we will set below.  In terms of
$a$ and $a_D$ the Lagrangian reads:
\eqn\lagaad{
\inv{32\pi}\Im \left\lbrack
(da)(d\bar{a}_D) + \left({da_D \over da}\right)
(F^2+iF\widetilde{F}) \right\rbrack.}
The coordinate $a$ lives on a cylinder since
\eqn\coorcy{\phi_2 \sim \phi_2 + {1 \over R_5}.}
This identification follows {}from large gauge transformations in five
dimensions which wind around the circle whose radius is $R_5$.  In five
and six dimensions the moduli space is $\bR^+ = \bR/\IZ_2$.  Therefore,
a good global coordinate on the moduli space is \nekrasov
\eqn\deful{u = 2\cosh(2\pi R_5 \sin\varphi\phi_1 + 2\pi i R_5 \phi_2).}
so we find
\eqn\aadn{\eqalign{
a =& v_5\cosh^{-1} (u/2),\cr
a_D =& v_6\cosh^{-1}(u/2).\cr}}

In terms of the six dimensional variables $\Phi$ and $B_{56}$
\eqn\uda{u = 2\cosh (4\pi^2 R_5 R_6 \sin\varphi \Phi + i B_{56}).}
This agrees with the fact that the masses of winding states of the
string around $R_6$ and $R_5$ should be $a$ and $a_D$ on one hand and on
the other hand $2\pi R_6\Phi$ and $2\pi R_5\Phi$ (for $B_{56}=0$) since
$\Phi$ is the tension of the string.  Furthermore, for large $\Phi$ we
can interpret ${1 \over u} \approx e^{-4\pi^2 R_5 R_6 \sin\varphi \Phi -
i B_{56}}$ as an instanton factor.  The relevant instanton is a string
which wraps our torus.

The central charge formula can be derived by matching to our five
dimensional expressions and by using the symmetry $5 \leftrightarrow 6$
\eqn\mass{
Z = n_e a + n_m a_D - 2\pi i (P_5 v_5 - P_6 v_6).}
Here $P_5$ and $P_6$ are the
momenta around $R_5$ and $R_6$ measured in integer units.

The monodromy of \aadn\ around $u=\infty$ is
\eqn\monaad{\eqalign{
a \approx& v_5\log u
     \longrightarrow a + 2\pi i v_5,\cr
a_D \approx & v_6 \log u
   \longrightarrow a_D +2\pi i v_6.\cr}}
This agrees with \mass\ if we change
\eqn\nenm{\eqalign{
n_e \longrightarrow & n_e, \cr
n_m \longrightarrow & n_m, \cr
P_5 \longrightarrow & P_5 - n_e, \cr
P_6 \longrightarrow & P_6 + n_m, \cr}}
This non-trivial monodromy becomes trivial when we consider the action
on $\partial a \over \partial u$ and $\partial a_D \over \partial u$.
Therefore, it is trivial when reduced to $SL(2,\BZ)$---it does not act
on $\tau={\partial a_D \over \partial a}$.  Such non-trivial monodromies
which include $SL(2,\BZ)$ as a quotient were observed in
\swii.  We see that although the gauge charges $n_e$ and $n_m$
transform trivially, the global charges $P_5$ and $P_6$ mix with the
gauge charges.

We have seen in the previous section that there is one additional
singularity at $\phi = \inv{\sqrt{2}R_6}$ in 5D.  Similarly, in 4D there
should thus be two additional singularities---one at $a =
\inv{\sqrt{2}R_6}$ and one at $a_D = \inv{\sqrt{2}R_5}$.  The
$SL(2,\BZ)$ monodromy around the first one is $T=\pmatrix{1&1\cr0&1}$
and around the second one is $S^{-1}T S=\pmatrix{1&0\cr-1&1}$. Since the
$SL(2,\BZ)$ monodromy around $u=\infty$ is trivial, we learn that the
monodromy around the remaining singularities is $\left(S^{-1}T S
T\right)^{-1} = ST$ which is indeed what we expect {}from an $E_8$
singularity.

In five dimensions we identified $n_e=L_6$.  It is clear that in four
dimensions we should identify $n_m=L_5$.  The monodromy \nenm\ changes
the quantum numbers of the particles at these singularities.
Therefore, if a particle with $(P_6,L_6,P_5,L_5) $ exists, there
should also be a particle with $(P_6-L_5,L_6,P_5+L_6,L_5)$ (note that
equation \sumstates\ is invariant under such a change in the charges).
Therefore, the particle at one singularity has
$(P_6=1,L_6=1,P_5=n,L_5=0) $ for arbitrary $n$.  All these particles
should exist and are permuted under the monodromy \nenm.  This fact
has a simple interpretation.  Upon reduction to five dimensions we
found a particle with $p_6=L_6=1$.  As a five-dimensional particle it
has arbitrary momentum.  In particular, its $P_5$ must be arbitrary.
A similar discussion applies to the particle at the other singularity
after a substitution $5 \leftrightarrow 6$ in all these equations.

%------------------------------------------------------------------%
%  The elliptic curve                                              %
%------------------------------------------------------------------%
\subsec{The elliptic curve}

To determine the elliptic curve we pick an auxiliary parameter $u$ on
the moduli space.  We assume that large $u$ corresponds to large tension
of the strings.

The elliptic curve
\eqn\ellip{
y^2 = x^3 - f(u,w_1\dots w_8,\sigma) x - g(u,w_1\dots w_8,\sigma)}
is determined by the following three observations \gantc:

\item{1.} The degrees of $f$ and $g$ in $u$ are 4 and 6 respectively.
This follows {}from the known behavior at large $u$.

\item{2.} The points $w_k$ equal the integrals of the meromorphic
two-form
\eqn\mero{\Omega = {du\wdg dx \over y}}
over eight 2-cycles of the total $x$-$u$ space which generate
an $E_8$ lattice.

\item{3.} The transformation:
\eqn\transf{y\rightarrow\alpha^3 y,\qquad x\rightarrow\alpha^2 x,
\qquad u\rightarrow\alpha u + \beta}
does not change the two-form $\Omega$ and hence does not change
the prepotential.

\item{4.} {}From the large $|u|$ behavior it follows that
\eqn\leadfg{\eqalign{
f(u) =& \inv{256 {v_5}^4} g_2(\sigma) u^4 + \ol{u^3},\cr
g(u) =& \inv{2048 {v_5}^6}g_3(\sigma) u^6 + \ol{u^5}.\cr
}}
where
\eqn\gg{\eqalign{
\inv{4}g_2(\sigma) =& 15 \pi^{-4}\sum_{m,n\in \IZ_{\ne
0}}\inv{(m\sigma+n)^4} \cr
\inv{4}g_3(\sigma) =& 35 \pi^{-6}\sum_{m,n\in \IZ_{\ne
0}}\inv{(m\sigma+n)^6} .\cr}}

These four requirements completely determine $f$ and $g$ up to the
transformation \transf.  The resulting functions are still in an
implicit form, though.  The explicit formulas are described in Appendix A.

We give below the formula for the curve corresponding to unbroken $E_8$:
\eqn\unbrk{
y^2 = x^3 - \inv{256 {v_5}^4} g_2(\sigma) u^4 x -
 (\a u^5 + \inv{2048 {v_5}^6} g_3(\sigma) u^6).}
The coefficient of $u^5$ is arbitrary.

\subsec{$d=4$, $N=2$, $SU(2)$ with $N_f=4$ and its $SL(2,\IZ)$ duality}

For generic Wilson lines, $w_i$, in \ellip\ the $E_8$ singularity can
split to ten singularities.  Adding to these the two singularities
discussed above (at $a={1 \over \sqrt{2} R_6}$ and at $a_D={1 \over
\sqrt{2} R_5}$) there are 12 singularities.  By tuning the Wilson lines
to a special value we can group them into two groups of six singularities
each.  In the notation introduced in the next section these two
singularities are of $D_4$ type.  In terms of four dimensional field
theory, the low energy theory near any of these singularities is $SU(2)$
with $N_f=4$ flavors.  A similar construction in the context of F-theory
compactification to eight dimensions appeared in \refs{\Sen, \bds}.

The $N=2$, $SU(2)$ with $N_f=4$ field theory is very special \swii.  It
is a finite quantum field theory with a dimensionless coupling constant
$\tau$.  It was argued in \swii\ that this theory exhibits $SL(2,\IZ)$
duality---the theory with the parameter $\tau$ is the same as the
theory with the parameter $\tau +1$ or $-{1 \over \tau}$.  This duality
acts in a non-trivial way on the spectrum of the theory.  Under $ \tau
\rightarrow \tau +1$ monopoles transform into dyons and under $\tau
\rightarrow -{1 \over \tau}$ electrons are interchanged with monopoles.
This means that $SL(2,\IZ)$ acts on the global $Spin(8)$ symmetry as
triality.  The evidence for this duality presented in \swii\ was based
on analyzing some of the spectrum and by examining the family of elliptic
curves which determine the effective gauge coupling $\tau_{eff}(u)$ on
the moduli space.

We now get a new understanding of this duality {}from our six-dimensional
viewpoint.    We are going to assume (as
everywhere in this paper) that 
the six-dimensional $E_8$ theory exists.  For large $|u|$ we see {}from
\newlag, \aad, and \lagaad\ that the effective coupling constant of the
$U(1)$ theory is given by the complex structure of the torus we
compactify on
\eqn\tauissig{\tau=\sigma.}
Since the $SU(2)$ gauge theory with $N_f=4$ is finite, this value of
$\tau$ is not corrected in the quantum theory and therefore \tauissig\
is true everywhere in the moduli space.

Consider now labeling the torus by $\sigma'$ which differs {}from
$\sigma$ by an $SL(2,\IZ)$ transformation.  Clearly, this should not
affect the effective theory.  But {}from \tauissig\ it is clear that the
value of $\tau$ of the $SU(2)$ theories is transformed by $SL(2,\IZ)$.
Therefore, the $SL(2,\IZ)$ of the torus which is a ``geometric
symmetry'' induces the $SL(2,\IZ)$ duality symmetry of the four
dimensional field theory.

This $SL(2,\IZ)$ duality has already been related to the $SL(2,\IZ)$
action on $T^2$ in \FHSV.  There, string compactification to four
dimensions was studied and the finite $N=2$ $SU(2)$ gauge theory
appeared in the low energy approximation.  The novelty of our
construction is that we use fewer degrees of freedom---a six-dimensional
field theory rather than string theory. 

Our understanding of the duality of the four dimensional, finite $N=2$
theory is similar to the understanding of the duality of the four
dimensional $N=4$ theory presented in \wittenII.
In both cases a six dimensional field theory at a non-trivial fixed
point of the renormalization group is compactified on a two torus with
complex structure $\sigma$ to four dimensions to yield a four
dimensional field theory with $\tau=\sigma$.  The duality of the four
dimensional theory is then understood as a consequence of the
$SL(2,\IZ)$ of the torus.

%==================================================================%
%  Singularities                                                   %
%==================================================================%
\newsec{Singularities in four dimensions}

In this section we examine what happens to the field theories of
\refs{\fived, \morsei} upon compactification to four dimensions.
Of course,  they can be obtained as compactifications of the
six-dimensional theory on $\bT^2$ as $R_6 \rightarrow 0$.

All these theories have a moduli space which is topologically the
complex $u$-plane.  The coupling constant of the photon is determined by
a torus fibered over this plane \swi.  This torus also determines the
metric on the moduli space.  The singularities in such a setup were
analyzed by Kodaira \kodaira.  All of them have realizations
in gauge theories:

\item{1.} $A_n$ ($n=0,1...$) singularities (corresponding to Kodaira's
I$_{n+1}$).  The monodromy around the
singularity is $T^{n+1}=\pmatrix{1&n+1\cr0&1}$ and the gauge coupling
behaves like
\eqn\uonemon{\tau = {n+1\over 2\pi i} \log z + \CO(1).}
These theories arise in $U(1)$ gauge theories with $n+1$ electrons.  The
global symmetry of the theory is $SU(n+1)$ and the Higgs branch is the
moduli space of $SU(n+1)$ instantons \swii.  All these theories are IR
free.

\item{2.} $D_n$ ($n=4,5...$) singularities (corresponding to Kodaira's
I$^*_{n-4}$).  The monodromy around the
singularity is $PT^{n-4}=\pmatrix{-1&-n+4\cr0&-1}$: the gauge coupling
behaves like
\eqn\sutwomo{\tau = {n-4\over 2\pi i} \log z + \CO(1),}
and the periods behave like
\eqn\sutwoper{\eqalign{
  a&=\sqrt z+\CO(1)\cr
  a_D&={n-4\over\pi i}\,a\log a+\CO(1).\cr}}
These theories arise in $SU(2)$ gauge theories with $n$ quark
hypermultiplets.  The global symmetry of the theory is $SO(2n)$ and the
Higgs branch is the moduli space of $SO(2n)$ instantons \swii.  For
$n>4$ these theories are IR free and for $n=4$ they are finite conformal
field theories for all values of the bare coupling constant $\tau$.

\item{3.} $H_n$ ($n=0,1,2$) singularities (corresponding to Kodaira's II,
III, IV). $\tau$ is determined by the
family of elliptic curves
\eqn\hncurves{\eqalign{y^2&= x^3 -z \qquad {\rm for} ~ H_0 \cr
y^2&= x^3 -zx \qquad {\rm for} ~ H_1 \cr
y^2&= x^3 - z^2 \qquad {\rm for} ~ H_2 .\cr}}
The monodromies around these singularities are
\eqn\hnmono{
\eqalign{(ST)^{-1}&=\pmatrix{1&1\cr-1&0} \qquad {\rm for} ~ H_0 \cr
S^{-1}&=\pmatrix{0&1\cr-1&0} \qquad {\rm for} ~ H_1 \cr
(ST)^{-2}&=\pmatrix{0&1\cr-1&-1} \qquad {\rm for} ~ H_2 .\cr}}
They arise in $SU(2)$ gauge theories with $n+1$ quark flavors by tuning
the quark mass parameter to a special value \apsw\ (the singularity
$H_0$ was first found in $SU(3)$ gauge theories in \ardou).
The global symmetry of the $H_n$ theory is $SU(n+1)$ (no symmetry for
$H_0$) and a Higgs branch isomorphic to the moduli space of $SU(n+1)$
instantons (no Higgs branch for $H_0$).  All of these are non-trivial
conformal field theories.  As \hncurves\ are deformed, they can
split to $n+2$ $A_0$ singularities.

\item{4.} $E_n$ ($n=6,7,8$) singularities (corresponding to Kodaira's
IV$^*$, III$^*$, II$^*$).  $\tau$ is determined by the
family of elliptic curves
\eqn\encurves{\eqalign{y^2&= x^3 + z^4 \qquad {\rm for} ~ E_6 \cr
y^2&= x^3 + xz^3 \qquad {\rm for} ~ E_7 \cr
y^2&= x^3 + z^5 \qquad {\rm for} ~ E_8 .\cr}}
The monodromies around these singularities are
\eqn\enmono{\eqalign{
(ST)^2&=\pmatrix{-1&-1\cr1&0} \qquad {\rm for} ~ E_6 \cr
S&=\pmatrix{0&1\cr-1&0} \qquad {\rm for} ~ E_7 \cr
ST&=\pmatrix{0&-1\cr1&1} \qquad {\rm for} ~ E_8 .\cr}}
As we compactify the $E_n$ ($n=6,7,8$) five-dimensional theories on a
circle, the $E_n$ symmetry must be present in the four-dimensional
theory and therefore we should find these theories.  (These $E_n$ cases
were recently analyzed in \refs{\minnem, \lerwar, \minnemtwo}.)
The non-trivial
conformal field theories associated with these have $E_n$ global
symmetry and the Higgs branch is the moduli space of $E_n$ instantons.

In this classification we did not include what can be called $D_n$ for
$n=0,1,2,3$; i.e.\ the results of $SU(2)$ with $n$ quark flavors.
These theories are UV free and are strongly coupled at long distance.
{}From the solution of these field theories \refs{\swi, \swii} we learn
that the singularities are
\eqn\lowdna{\eqalign{
A_0+A_3 &\qquad {\rm for}~ n=3 \cr
A_1+A_1  &\qquad {\rm for}~ n=2 \cr
A_0+A_0+A_0  &\qquad {\rm for}~ n=1 \cr
A_0+A_0  &\qquad {\rm for} ~n=0 .}}

As a preliminary for matching the five-dimensional theories with the
four-dimensional ones, let us examine the behavior for large $|u|$.
The metric on the moduli space of the five-dimensional theory there is
$(ds)^2=(t_0 + {c\over \sqrt{2}}\phi) \,d\phi \,d\phi$.  The dimensional
reduction leads to a factor of $R_5$ in this metric and to another
compact scalar $\theta= R_5 A_5\sim \theta + 2\pi$ with metric
\eqn\fourdmet{(ds)^2=R_5({16\pi^2 \over {g_0}^2}+ {c\over \sqrt{2}}\phi)
 \,d\phi \,d\phi+
{1 \over 8\pi^2 R_5}({16\pi^2\over {g_0}^2} + {c\over \sqrt{2}}\phi)
 \,d\theta \,d\theta.}

To cast it in $N=2$ superspace we define
\eqn\fourddef{\eqalign{a&=\phi + {i\theta\over 2\sqrt{2}\pi R_5}
= {1 \over 2\sqrt{2}\pi R_5} \log u \cr
a_D&=2\pi i R_5 ({8\pi \over {g_0}^2} a + {c\over 4\sqrt{2}\pi} a^2)
\cr}}
As in \monaad, we can work out the monodromy under $u \rightarrow
e^{2\pi i} u$.  This time it is non-trivial even in $SL(2,\BZ)$---it is
\eqn\moninfty{\CM=T^{c/2}.}
In the $A_n$ case (i.e., the $U(1)$ theories in five dimensions), $c$
can be
odd and this calculation is another indication that the $U(1)$ theory
cannot be considered in isolation.  However, if the $U(1)$ theory is
embedded in another theory, the change in monodromy $T^{c/2}
(T^{-c/2})^{-1}=T^c$ can be measured in that theory, and is sensible.
In the $D_n$ and $E_n$ cases, $c$ is even and this complication does not
arise.

We can now consider the various five-dimensional theories:

\item{1.} $U(1)$ with $n$ electrons.  The moduli space $\bR$ becomes
$\bR \times \bS^1$ in four dimensions where the $\bS^1$ originates
{}from the $A_5$ component of the gauge field.  Classically, the metric
is as in \fourdmet\ with $c=0$
and there are $n$ massless electrons at $\phi=0$.  The one loop
correction to the metric includes contributions of the massive
Kaluza-Klein modes in the electrons and is given by a simple integral.
Even without examining it explicitly, we can easily extract some of
its properties.  For large $|\phi|$ the metric and the monodromies are
determined as in \fourdmet\ and \moninfty.  Therefore, the monodromy
around $\phi=0$ is $T^{-n}$.  This is consistent with the $A_{n-1}$
singularity which is expected there.  Indeed, as these theories are IR
free both in five and in four dimensions, we do not expect the long
distance behavior to change the tree level spectrum.

\item{2.} $SU(2)$ with $n$ quarks. The moduli space $\bR/\BZ_2$ becomes
$(\bR \times \bS^1)/\BZ_2$ in four dimensions where the $\bS^1$
originates {}from the $A_5$ component of the gauge field and $\BZ_2$
{}from the Weyl group of $SU(2)$.
Classically, the metric is as in \fourdmet\ with $c=0$.  There are two
special points.  The obvious one is at $\phi=\theta=0$: there is a
four-dimensional $SU(2)$ theory with $n$ quarks.  Somewhat less obvious
is the point $\phi=0$, $\theta=\pi$, where there is an $SU(2)$ theory
with no quarks.  In terms of the variable $u$ of \deful\ these two
singularities are at $u=\pm 1$.  Quantum mechanically this picture
changes.  First, at one loop the metric is corrected.  The monodromy at
large $u$ is determined by \moninfty\ to be $T^{8-n}$.  It is equal to
the product of the monodromy around $\phi=\theta=0$ which is $PT^{4-n}$
and the monodromy around $\phi=0$, $\theta=\pi$ which is $PT^4$ (where
$P=\pmatrix{-1&0\cr0&-1} \in SL(2,\BZ)$).  Non-perturbatively, the theory
at $\phi=0$, $\theta=\pi$ is always strongly coupled and the theory at
$\phi=\theta=0$ is strongly coupled for $n<4$.  Using \lowdna, we
determine the singularity structure in four dimensions for the reduction
of the various five-dimensional theories

\eqn\lowdnn{\eqalign{D_n & \rightarrow D_n +A_0+A_0 \qquad {\rm for}\
n \ge 4 \cr
D_3 &\rightarrow A_0+A_0+A_0+ A_3 \cr
D_2 &\rightarrow  A_0+A_0+A_1+A_1 \cr
D_1 &\rightarrow A_0+A_0+A_0+A_0+A_0\cr
D_0 &\rightarrow  A_0+A_0+A_0+A_0 \cr
\tilde D_0 &\rightarrow  A_0+A_0+A_0+A_0  \cr}}
where in the last one we used the fact that the long distance Lagrangian
is the same as for $D_0$---they differ only in a $\IZ_2$ theta-like
angle \dkv.

\item{3.}  The non-trivial theories $E_n$.  We have already seen in the
previous section that the $E_8$ singularity splits
\eqn\eeights{E_8 \rightarrow E_8+A_0.}
Similarly,
\eqn\esevs{\eqalign{E_7 &\rightarrow E_7+A_0\cr
E_6 &\rightarrow E_6+A_0.}}
One way to derive this is by perturbing the
$E_8$ five-dimensional theory and examining the consequences in four
dimensions.  We can continue to flow down and determine what happens to
the other non-trivial five-dimensional theories.  $E_5$ has $SO(10)$
global symmetry and a Higgs branch which is the moduli space of $SO(10)$
instantons.  There is only one theory in four dimensions with this Higgs
branch and a one-dimensional Coulomb branch.  This is the $SU(2)$ theory
with $n=5$ whose singularity is $D_5$.  Matching with the flow {}from
$E_6$ in five dimensions we conclude
\eqn\efivefl{E_5 \rightarrow D_5+A_0 .}
Since this theory is IR free, continuing to flow down is easy
\eqn\lowen{\eqalign{
E_4 &\rightarrow A_4 +A_0 +A_0 \cr
E_3 &\rightarrow A_2 +A_1 +A_0 \cr
E_2 &\rightarrow A_1 +A_0+A_0 +A_0 \cr
E_1 &\rightarrow A_1 +A_0 +A_0 \cr
\tilde E_1 &\rightarrow A_0 +A_0 + A_0 +A_0 \cr
E_0 &\rightarrow A_0 +A_0 +A_0 \cr
}}

\item{4.}  Special values of the parameters.  Here we can find the
$H_n$ singularities by starting with $D_{n+1}$ in five dimensions with
equal non-zero masses and tuning it to an appropriate value.

\newsec{Applications to string theory}

\subsec{F-theory models in eight dimensions}

In our first application to string theory of our analysis of the
compactification to dimension four we encounter F-theory again, but in a
different context.  An F-theory model in eight dimensions can be
regarded as a compactification of the type IIB string on a two-manifold
$B$ with some background D7-branes (generically, with $24$ distinct
D7-branes).  The relation to our point of view is provided by the use of
a D3-brane probe \bds.  The field theory on a D3-brane probe has $B$ as
its moduli space.  This can be seen quite explicitly as follows: on the
one hand, the field theory is described in terms of the gauge coupling
$\tau$ which depends on a family of elliptic curves; on the other hand,
the same function $\tau$ is used to build the F-theory model, where it
plays the role of the complexified dilaton on which $SL(2,\BZ)$ acts.
The possible singularities in the family of elliptic curves---which
correspond to possible ways that the D7-branes coalesce---are again
described by Kodaira's classification.  In the notation we use here,
these are $A_n$ ($n\ge0$), $D_n$ ($n\ge4$), $E_n$ ($n=6,7,8$) and $H_n$
($n=0,1,2$).  The $A_n$ singularities correspond to the enhanced
$SU(n+1)$ gauge symmetry {}from coalescing $n+1$ D7-branes
\refs{\nclp, \bound, \BSV}, the $D_n$ singularities correspond to the enhanced
$SO(2n)$ gauge symmetry from coalescing $n$ D7-branes and an orientifold
plane \refs{\nclp, \Sen, \bds} and the $E_n$ singularities occur at
strong coupling, as suggested in \rDM.

The splitting of singularities when compactifying {}from five to four
dimensions can be seen directly in this application.  The
nine-dimensional type I$'$ compactification generically (on an open set
in its moduli space) has two orientifold planes and 16
D8-branes.\foot{This is only one of three possible generic behaviors,
each being valid on an appropriate open set in moduli \morsei.  The
other two are (i) two orientifold planes, one of which is strongly
coupled (giving an $E_0$ theory rather than a $D_0$ theory), and 17
D8-branes, and (ii) two strongly coupled orientifold planes (giving
$E_0$ theories) and 18 D8-branes.  All three possibilities lead to 24
$A_0$ points after further compactification.} The corresponding
five-dimensional
field theory content is 2 $D_0$ points and 16 $A_0$ points; in four
dimensions, the $D_0$'s each split to 4 $A_0$'s, giving a total of 24
$A_0$'s, i.e., 24 D7-branes in F-theory.

Similar analyses can be made for non-generic singularities; to do so, we
need to discuss gauge symmetry enhancement in these models.  The
possible ways in which the D8-branes can coalesce can be studied
directly by considering the heterotic theories which are dual to the
type I$'$ theories---the coalescing D8-branes correspond to gauge
symmetry enhancement. In the heterotic interpretation, the root lattice
of the gauge group must be embedded in the even unimodular lattice of
signature $(1,17)$, and for any such embedding, appropriate expectation
values for the Wilson lines can be chosen to ensure that the gauge group
is the one corresponding to such a sublattice.

When we further compactify on a circle, we must look for embeddings of
the root lattice of the gauge group into the even unimodular lattice of
signature $(2,18)$.  This can be seen either {}from the heterotic point of
view, or {}from the F-theory point of view.  In the F-theory version, we
must find a K3 surface which has an elliptic fibration with a section,
and some rational curves (contained in fibers of the fibration) whose
intersection matrix reproduces the root system in question.  Splitting
off the classes of the base and fiber of the fibration {}from the
cohomology lattice of the K3 leaves an even unimodular lattice of
signature $(2,18)$ into which the root lattice must be embedded.  Thanks
to the global Torelli theorem for K3 surfaces, for every such embedding
there is a K3 surface which realizes it, so there is an F-theory model
with the specified gauge symmetry group.  It can always be realized by
writing a Weierstrass equation in which the corresponding K3 surface has
singular points of precisely the types specified by the root system.
(The rational curves arise upon resolving those singularities.)

Note that the $(1,17)$ lattice embeds into the $(2,18)$ one, so any
gauge group realized in nine dimensions is also realized in eight
dimensions, consistent with the splitting of singularities we have
discussed.  There are effective techniques in the mathematics literature
for determining whether a given root lattice has an embedding into the
$(1,17)$ or $(2,18)$ lattice \nikulin, and for determining the number of
inequivalent embeddings \refs{\nikulin, \mirmor}.

Although for any specified embedding of a root system into the $(2,18)$
lattice, there is an F-theory realization with the corresponding gauge
symmetry, this data does not completely determine the content of the
four-dimensional field theory on the probe: a model with an $H_n$ point
is indistinguishable in this regard {}from a model in which the $H_n$
point is replaced by an $A_n$ point plus an $A_0$ point.  For this
reason, some care must be taken in applying lattice-embedding techniques
to produce explicit models.  For example, to determine if there exists a
model with two $E_8$ points and an $A_2$ point, we note that the
corresponding gauge group $E_8\times E_8\times SU(3)$ does occur on
F-theory models \rDM.  (An explicit model was constructed in \rDM, but this
fact can also be checked using lattice embedding techniques---the lattice
embedding turns out to be unique.)
The potential ambiguity between $A_2$ and $H_2$ points leads us to
pursue a more explicit method.  If we begin with the Weierstrass
equation which describes an F-theory model with gauge group containing
$E_8\times E_8$, given as in
\MVII\ by
\eqn\eeightf{y^2=x^3+\alpha xz^4+z^5+\beta z^6+z^7,}
and compute its discriminant
\eqn\eeightd{-z^{10}\left(27z^4+54\beta
z^3+(4\alpha^3+27\beta^2+54)z^2+54\beta z+27\right).}
then the $E_8$ singularities are at $z=0$ and at $z=\infty$. If there is
in addition an $H_2$ point or an $A_2$ point we can change coordinates
to put it at $z=1$; we then need a triple zero of the discriminant at
that point. The only way to achieve that is to set
$\alpha=0$ and $\beta=-2$,  leaving us with the equation
\eqn\eeighth{y^2=x^3+z^5(z-1)^2,}
which has an $H_2$ point at $z=1$, not an $A_2$ point.

In \morsei\ it was observed that there is a type I$'$ model with an
$E_0$ theory at each orientifold plane, and an $A_{17}$ at another
point, leading to $SU(18)/\BZ_3$ enhanced gauge symmetry.  (In fact, the
existence of this theory was a crucial step in establishing the
existence of the $E_0$ field theories.)  In four dimensions, each $E_0$
point is expected to split to 3 $A_0$ points, leaving us with an
F-theory model which (generically) has 6 $A_0$ points and one $A_{17}$
point.  Using lattice embedding techniques, the existence an F-theory
model with $SU(18)/\BZ_3$ in its gauge group can easily be established.
(The quotient by $\BZ_3$ corresponds in lattice-theoretic terms to the
fact that the cokernel of the embedding map contains $\BZ_3$ as its
torsion subgroup; this is directly related to the need to use an
exceptional modular invariant to construct the corresponding heterotic
theory \morsei.  This phenomenon was also apparent in the
$Spin(32)/\BZ_2$ models constructed in \aspgr.)  To see that we actually
get 6 $A_0$ points rather than some mixture of $H_0$ and $A_0$ points,
we write an explicit model, with an equation of the form\foot{We should
point out that the version of Tate's algorithm \tate\ which was
formulated in \bikmsv\ requires some care in its application to this
case: because the order of vanishing of the discriminant is so large,
the equation cannot be manipulated into the form specified in table 2 of
\bikmsv\ without introducing meromorphic changes of coordinates on the
base $B$.  In fact, when the discriminant of this equation is
calculated, there are three miraculous cancellations of terms which
allow the discriminant to vanish to order $18$ at $z=0$.}
\eqn\esueighteen{y^2 + (a z^2 + b z + c) xy +  z^6 y = x^3,}
in which the $SU(18)$ occurs
at $z=0$. The discriminant of this cubic equation is
\eqn\eighteend{-{1\over16}z^{18}\left(27z^6-(az^2+bz+c)^3\right),}
so the other singularities are at the zeros of $27z^6 - (az^2+bz+c)^3$,
which generically has six distinct zeros (each giving an $A_0$ point).
This confirms both the existence of the $E_0$ theory in five dimensions,
and its splitting into 3 $A_0$ points in four dimensions.

Similar remarks apply to the $Spin(34)$ model of \morsei.

\subsec{IIA theory compactified on a Calabi--Yau threefold}

In our second application to string theory, we have five-dimensional
models obtained by compactifying M-theory on a singular Calabi--Yau
threefold, {}from which we can produce four-dimensional models by further
compactification on a circle.  In an appropriate domain in the moduli
space these four-dimensional models can be interpreted as IIA string
theory compactified on Calabi--Yau threefolds.  The modifications to the
moduli space arise {}from two sources: (1) the moduli includes a new
$2$-form in the NS-NS sector---the $B$-field---arising {}from zero-modes
of the M-theory $3$-form integrated along the circle, and (2) the moduli
space
is corrected by worldsheet instantons, which can be interpreted as the
worldvolume of the M-theory two-brane wrapping $S^1\times\Sigma^{(i)}$ for
surfaces $\Sigma^{(i)}$ in the Calabi--Yau manifold.  The first modification
complexifies the scalar in the vector multiplet, and the second is
responsible for various quantum effects such as the splitting of
singularities.  We explore this latter point in detail below.

Let us first analyze these quantum effects on general grounds.  Let
$A_5^{(i)}$ be the area of the surface $\Sigma^{(i)}$, measured in
M-theory units.
The relationship between
the M-theory scales and the type IIA scale and coupling then imply that
\eqn\fourv{ R_5A_5^{(i)} \sim T^{(i)},}
where we have denoted the area of $\Sigma^{(i)}$ in type IIA units by
$T^{(i)}$.  Note that $T^{(i)}$ appears in a vector multiplet in the
four dimensional field theory and hence the metric on the Kahler moduli
space can depend on it.  Another scalar made out of $R_5$ and the volume
of the Calabi-Yau space in M theory units, $S$, is in a hypermultiplet
and therefore cannot affect this metric.

We normalize the action so that worldsheet instantons of charge $n$
contribute $e^{-2\pi nT^{(i)}}$ (multiplied by a phase); these can be
interpreted as the wrapping of the M-theory two-brane worldvolume around
$S^1\times\Sigma^{(i)}$ (consistent with \fourv).  We let $J$ and $B$
denote the K\"ahler form and $B$-field, and introduce the notation
\eqn\enotq{q^\sigma=e^{2\pi i\sigma\cdot(B+iJ)}}
for the instanton contributions (where $\sigma$ is the homology class of
$\Sigma$) so that $q^{\sigma^{(1)}}$, \dots, $q^{\sigma^{(k)}}$ serve as
coordinates on the K\"ahler moduli space.

If we pass to the field theory limit in five dimensions before
compactifying, only a subset of the instantons will be available to
correct the field theory correlation functions.  In the K\"ahler moduli
space, this corresponds to setting $q^{\sigma^{(j)}}=0$ for any $j$ for
which $A_5^{(j)}$ decouples {}from the field theory.  Typically this
restricts the computation to the boundary of the K\"ahler moduli space,
along which the Calabi--Yau manifold is singular. Techniques for
computing in such limits were developed in \small.

In the $A_n$ case, the worldsheet instanton sum associated to a flop
transition was analyzed in \refs{\phases, \small}.  Performing a flop on
$n+1$ rational curves (all {}from the same homology class) alters the
intersection numbers of divisors on the Calabi--Yau threefold, which is
why the five-dimensional gauge coupling is only a piecewise linear
function.  However, the corrections from wrapped two-branes modify the
singularity so that the coupling in four dimensions has a pole rather than
a discontinuous derivative \mandf.

A typical correlation function is given by the intersection number
$H_1\cdot H_2\cdot H_3$ in the M-theory context.  (The gauge coupling
can be determined {}from the behavior of $H\cdot H\cdot H$ for
appropriate divisors $H$ \refs{\ccdf, \fkm, \fms, \mandf, \morsei}.)  If
$\gamma$ is the common homology class of the $n+1$ rational curves being
contracted, this correlation function is corrected by instantons to the
following value in the four-dimensional theory
\eqn\couplinst{\langle H_1\,H_2\,H_3\rangle=
H_1\cdot H_2\cdot H_3 + (n+1)\,{q^\gamma\over1-q^\gamma}\,
(\gamma\cdot H_1) (\gamma\cdot H_2) (\gamma\cdot H_3),}
valid for small $|q^\gamma|$.
(We have
suppressed all instanton corrections other than the ones related to
$\gamma$, which is the only relevant class in the field theory.)  The
expression \couplinst\ for the coupling
can be analytically continued past the singularity at $q^\gamma=1$,
by employing the identity
\eqn\econtinue{{q^\gamma\over1-q^\gamma}=-1-{q^{-\gamma}\over
1-q^{-\gamma}}}
to yield
\eqn\couplbis{\eqalign{\langle H_1\,H_2\,H_3\rangle=
&\left(H_1\cdot H_2\cdot H_3 - (n+1)
(\gamma\cdot H_1) (\gamma\cdot H_2) (\gamma\cdot H_3)
\right)\cr
&+ (n+1)\,{q^{-\gamma}\over1-q^{-\gamma}}\,
(-\gamma\cdot H_1) (-\gamma\cdot H_2) (-\gamma\cdot H_3).}}
The first term in \couplbis\ is the M-theory coupling (i.e., the
classical intersection product) on the flopped
Calabi--Yau manifold
\eqn\flopcoup{\widehat H_1\cdot\widehat H_2\cdot\widehat H_3=
H_1\cdot H_2\cdot H_3 - (n+1)
(\gamma\cdot H_1) (\gamma\cdot H_2) (\gamma\cdot H_3),}
and the entire expression \couplbis\ is seen to be precisely
the instanton-corrected value for
$\langle\widehat H_1\,\widehat H_2\,\widehat H_3\rangle$ (bearing in
mind that the homology class of the instanton changes sign during the
flop).

In the $D_n$ case, the instantons shrinking to zero size are a bit more
complicated.  Let $2\gamma$ be the homology class of the fiber of the
ruled surface $S$ which is shrinking, so that $\gamma$ is the class of
any of the components of the singular fibers in that ruling.  We have
$n$ singular fibers in the ruling, each with $2$ components, so there
are a total of $2n$ instantons in class $\gamma$.  On the other hand, in
the homology class $2\gamma$ there is an entire $\CP1$ of shrinking
rational curves which according to the calculations of \rKMP\ contribute an
instanton number of $-2$.  There are no instantons in any other
multiples of $\gamma$.  Thus, the entire instanton correction to
$H_1\cdot H_2\cdot H_3$ due to the holomorphic curves which shrink to zero
size at the $D_n$ point is
\eqn\Dinst{\eqalign{
&2n{q^\gamma\over1-q^\gamma}\, (\gamma\cdot H_1) (\gamma\cdot H_2)
(\gamma\cdot H_3) + -2{q^{2\gamma}\over1-q^{2\gamma}}\,(2\gamma\cdot
H_1) (2\gamma\cdot H_2)  (2\gamma\cdot H_3) \cr
&=
\left(2n{q^\gamma\over1-q^\gamma}-2{8q^{2\gamma}\over
1-q^{2\gamma}}\right)
(\gamma\cdot H_1) (\gamma\cdot H_2) (\gamma\cdot H_3)\cr
&=\left((2n-8){q^\gamma\over1-q^\gamma}-8{q^{\gamma}\over-1-q^{\gamma}}
\right)(\gamma\cdot H_1) (\gamma\cdot H_2) (\gamma\cdot H_3).}}
This computation clearly shows the splitting of the five-dimensional
singularity to two singularities in four dimensions, at $q^\gamma=\pm1$,
whose locations only differ by the value of the $\gamma$-component of
the $B$-field, which is $0$ in one case and $1/2$ in the other.  This
feature of $D_n$-type Calabi--Yau theories does not seem to have been
observed before.

As in
\morsei, the gauge coupling in five dimensions is determined by
$S\cdot S\cdot(\phi S+t_0H_0)$ where $H_0$ is a divisor meeting $S$ in a
section of the ruling; this is the second derivative of the prepotential
${\cal F}$ with respect to the parameter $\phi$ associated to $S$.  The
quantum corrected version of the {\it third}\/ derivative is given by
\eqn\thirdderiv{
{\partial^3{\cal F}\over\partial\phi^3}=\langle S\,S\,S\rangle
= (8-n) +
(-1)^3\left((2n-8){q^\gamma\over1-q^\gamma}-8{q^{\gamma}\over
-1-q^{\gamma}} \right),}
suppressing other instanton corrections.  We identify the period $a$ in
the field theory with the area of $\gamma$, (i.e., $a=-\phi$ since
$\gamma\cdot S=-1$), and the period $a_D$ as
\eqn\aD{a_D={\partial{\cal F}\over\partial a}=-{\partial{\cal
F}\over\partial \phi};}
we also have $q^\gamma=e^{-2\pi i\phi}$.  Then the leading order
behavior of $\partial^3{\cal F}/\partial\phi^3$ near $q^\gamma=1$ is
\eqn\leadthird{
{\partial^3{\cal F}\over\partial\phi^3}=-{2n-8\over2\pi i\phi}+\cdots,}
so the leading order behavior of $a_D$ is
\eqn\aDlead{a_D={2n-8\over2\pi i}a\log a+\cdots.}

The generator of the monodromy is described as in \rKMP\ by the {\it
elementary transformation}\/ associated to the ruled surface $S$
\refs{\burnsrap, \wilson}.  That transformation is a map on cohomology
\eqn\elttr{\rho(H)=H+(2\gamma\cdot H)S,}
which is a kind of reflection in the class $S$, mapping $S$ to $-S$.  It
thus acts on $\phi$ and on $a$ as multiplication by $-1$; combining with
\aDlead\ we see that the monodromy action agrees with \sutwoper, as
expected for a $D_n$ point.

This same elementary transformation generates monodromy about
$q^\gamma=-1$ as well, since $\phi\mapsto -\phi\pmod\BZ$ has fixed points
at both $0$ and $1/2$.  The monodromy at $1/2$ is exactly as expected
for a $D_0$ point, by the same computation.  We confirm in this way the
splitting of the $D_n$ point {}from dimension five into a $D_n$ and a
$D_0$ point in dimension four.

In this computation, we have suppressed instantons which wrap other
holomorphic curves in the surface which shrinks to zero size.  That is a valid
approximation if the base of the ruled surface is taken to be extremely
large---in this approximation, the quantum corrections to the
four-dimensional field theory are suppressed so we see the ``classical''
splitting into two singularities without the further ``quantum''
splitting of one or both of those singularities.  To see the full
description we should consider these other instantons.

Let $\Sigma^{(1)}$ be the base $\CP1$ of the ruled surface.
Then the five-dimensional gauge coupling for the $D_n$
theory is given by
\eqn\fivec{{1\over g_5^2} \sim A_5^{(1)},}
and the four-dimensional gauge coupling is
\eqn\fourc{{1\over g_4^2} \sim R_5A_5^{(1)} \sim T^{(1)}.}
We introduce the homology class $\eta$ of $\Sigma^{(1)}$ so that the
corresponding instanton contribution is $q^\eta$.

{}From the solution to the $D_0$
field theory, we predict a splitting of the singularity at $\gamma\cdot
B=1/2$ of the form
\eqn\splitu{u_+ - u_- \sim q^{\eta/2}.}
That is, including the
instantons wrapping the base $\CP1$ should modify the appropriate term
$-8q^\gamma/(-1-q^\gamma)$ {}from \Dinst\ to something of the
form
\eqn\Dinstbis{\eqalign{&
{-4q^\gamma+\ol{q^{\eta/2}}\over
\phantom{\lower6pt\hbox{x}}
-1-q^\gamma+ q^{\eta/2}
f(q^\gamma) + \ol{q^\eta}}
+{-4q^\gamma+\ol{q^{\eta/2}}\over
-1-q^\gamma- q^{\eta/2}f(q^\gamma)+\ol{q^\eta}
} \cr&=
{8q^\gamma(1+q^\gamma)+\ol{q^\eta}\over(1+q^\gamma)^2-q^\eta
f(q^\gamma)^2+\ol{q^{2\eta}}}} }
for some function $f(q^\gamma)$ which does not vanish at
$q^\gamma=-1$. (The terms $\ol{q^{\eta/2}}$ and $\ol{q^\eta}$ also depend
on $q^\gamma$.)

The denominator in \Dinstbis\ determines the location of a component of
the discriminant locus in the quantum-corrected (vector) moduli space of
the Calabi--Yau manifold.  There is another, related subset of the boundary
of the moduli space: the locus where $q^\eta=0$ and all
$q^{\sigma^{(j)}}=0$ for $\sigma^{(j)}\ne\eta,\gamma$; we have identified
this locus
with the weak coupling limit in the field theory.  It is a holomorphic
curve which is tangent to
the discriminant-locus component $(1+q^\gamma)^2-q^\eta
f(q^\gamma)^2+\ol{q^{2\eta}}=0$ defined by
\Dinstbis.

This qualitative feature of an asymptotically free $SU(2)$ theory---a
component of the discriminant locus which is tangent to a curve in the
boundary, due to
the quantum splitting of singularities---was presented by Kachru and
Vafa \kachvaf\ as evidence that they had correctly identified a
four-dimensional heterotic/type II dual pair \refs{\kachvaf, \FHSV}.  The
behavior near such tangent loci was subsequently investigated in
greater detail \refs{\KKLMV, \GHL}.  Here we observe that such a structure
will be a generic feature in Calabi--Yau moduli space any time a
rational ruled surface shrinks to zero size.  (Some related observations
have been made in \kkv.)  This can be seen even in models
with no manifest heterotic dual \refs{\BKKM, \kkv}.
The compatibility with heterotic/type II duality (if it is present)
becomes clear when one recalls that for Calabi--Yau models with a
heterotic dual, there is always a K3 fibration over a base $\Sigma$,
with the area of $\Sigma$ mapping to the heterotic coupling \asplouis;
moreover, a perturbative $SU(2)$ on the heterotic side corresponds
\aspgauge\ to a ruled surface on the Calabi--Yau threefold over the same
base $\Sigma$.

When $n\le3$, the singularity at $\gamma\cdot B=0$ should exhibit a similar
behavior.  That is, the other term
$(2n-8)q^\gamma/(1-q^\gamma)$ {}from \Dinst\ should be modified by instantons
to something of the approximate form
\eqn\Dinstbisoth{\eqalign{&
{(n-4)q^\gamma+\ol{q^{\eta/2}}\over
\phantom{\lower6pt\hbox{x}}
1-q^\gamma+ q^{\eta/2}
g(q^\gamma) + \ol{q^\eta}}
+{(n-4)q^\gamma+\ol{q^{\eta/2}}\over
1-q^\gamma- q^{\eta/2}g(q^\gamma)+\ol{q^\eta}
} \cr&=
{(2n-8)q^\gamma(1-q^\gamma)+\ol{q^\eta}\over(1-q^\gamma)^2-q^\eta
g(q^\gamma)^2+\ol{q^{2\eta}}}} }
for some function $g(q^\gamma)$ which does not vanish at
$q^\gamma=-1$.

To summarize our conclusions about the $D_n$ case: a rational ruled
surface shrinking to a curve is associated to two components of the
discriminant locus,\foot{There are two distinct components near the
weak-coupling limit, but they could be globally identified in some
examples.  In addition, it might happen that the map {}from the field
theory moduli space to the Calabi--Yau moduli space is many-to-one, and
these components could in principle have the same image.} one of which
is tangent to the weak coupling limit locus $q^\eta=0$,
$q^{\sigma^{(j)}}=0$.  The other component will also be tangent to that
locus if $n\le3$.  This entire structure is
associated to a {\it single}\/ $SU(2)$ factor of the gauge group.
The non-trivial dynamics in four dimensions are responsible for the
somewhat indirect way in which this $SU(2)$ manifests itself---it is more
clearly visible in five dimensions.

We turn now to the $E_n$ case, in which a surface $S$ shrinks to a
point.  We assume that we are in the generic situation in which the
image of $H^{1,1}(X)\to H^{1,1}(S)$ is one-dimensional.  The coefficient
$c$ which governs the five-dimensional field theory is calculated by the
intersection product $S\cdot S\cdot S$.  In four dimensions, this is
corrected by instantons to
\eqn\Einst{\langle S\,S\,S\rangle= (S\cdot S\cdot S)+\sum_{j=1}^\infty
N_j\,{j^3q^{j\gamma}\over1-q^{j\gamma}},}
where $\gamma$ is a homology class on $S$ such that $S\cdot \gamma=-1$,
and where $N_j$ is the instanton number associated with rational curves in
class
$j\gamma$.  In these $E_n$ cases, unlike the previous two, there will be
an infinite number of homology classes contributing to the instanton
sum, but the answer is universal for the $E_n$ theory in question
(depending only on $n$---of course there is also a sum for $\tilde
E_1$).  The number $N_1$ should be the number of ``lines'' on $S$, that
is, the number of $\CP1$'s whose ``degree'' $-S\cdot\gamma=c_1(S)\cdot
\gamma$ is $1$.  The numbers $N_j$ for $j>1$ have a less straightforward
interpretation, since the family of $\CP1$'s in such a class usually has
positive dimension.

In the case of $E_0$, the first several terms of this instanton expansion
\Einst\ were
calculated in \refs{\HKTY, \twoparamtwo}:
\eqn\Einstzero{\eqalign{
9&{}+3{3^3q^{3\gamma}\over1-q^{3\gamma}}
-6{6^3q^{6\gamma}\over1-q^{6\gamma}}
+27{9^3q^{9\gamma}\over1-q^{9\gamma}}
-192{12^3q^{12\gamma}\over1-q^{12\gamma}}\cr
&{}
+1695{15^3q^{15\gamma}\over1-q^{15\gamma}}
-17064{18^3q^{18\gamma}\over1-q^{18\gamma}}
+188454{21^3q^{21\gamma}\over1-q^{21\gamma}}
-\cdots,}}
and the meaning of the higher terms was explored in \twoparamtwo, where
a clear geometric interpretation was found for the first two nonzero terms
$N_3=3$, $N_6=-6$.  It was also noted there that many of the higher
terms, beginning with $N_6$, are negative.  In fact, the series appears
to alternate signs after the first term.

The splitting of $E_0$ into three singularities in four dimensions has a
clear interpretation {}from this point of view.  For the del Pezzo surface
$S=\CP2$ associated to the $E_0$ theory, $N_j$ must vanish unless $j$ is
divisible by $3$.  This is because $c_1(S)$ has intersection number $3$
with the generator of second homology.  Thus, the entire series
\Einstzero\ is a function of $q^{3\gamma}$, and it will have three poles
near the origin whose values differ by a cube root of unity.  At those
singularities, the value of the area $\gamma\cdot J$ will be shifted
away {}from zero \small\ and the $\gamma$-component of the $B$-field
$\gamma\cdot B$ will take one of three possible values $0$, $1/3$ or
$2/3$.\foot{For some examples, including the case discussed in \small,
there will be a three-to-one map {}from the field theory moduli space to
the Calabi--Yau moduli space, which identifies these three
singularities, so that only the value $\gamma\cdot B=0$ occurs.}

For the higher $E_n$'s some explicit calculations of \Einst\ were made
for the cases of $E_5$, $E_6$, $E_7$, $E_8$ in \kmv.  There is a
mysterious discrepancy \kmv\ between the first coefficient and the
number of lines in the case of $E_8$, but in the other cases the number
of lines corresponds to $N_1$ as expected.  These series also appear to
alternate signs after the first term.  It will be very interesting to study
these series further in
order to verify other aspects of our qualitative description.

\newsec{Compactification to three dimensions}

The compactification of all these theories to three dimensions is easily
analyzed either by using string duality as in \threedone\ or using field
theory methods as in \SWthreeD.  This section is mostly a review of
\refs{\threedone, \SWthreeD} in the notation of this paper.

In three dimensions the Coulomb branch is a hyper-K\"ahler manifold.  In
our case it is of real dimension four.  Three of the coordinates of this
manifold arise {}from the compactification of the six-dimensional
two-form.  The fourth one is dual to the three-dimensional vector.

The Coulomb branch for these theories is as follows.

\noindent
$U(1)$ gauge theory with $n$ electrons, $A_{n-1}$: The Coulomb branch
has an $A_{n-1}$ singularity.  In particular, for $n=1$ it is smooth.
The theories at the singularities are at non-trivial fixed points.

\noindent
$SU(2)$ gauge theory with $n$ quarks, $D_{n}$: The Coulomb branch
has a $D_n$ singularity.  In particular, for $n=0,1$ it is smooth, for
$n=2$ it has two $A_1$ singularities and for $n=3$ it has an $A_3$
singularity. The theories at the singularities are at non-trivial fixed
points.

In \SWthreeD\ these four-dimensional gauge theories were studied on
$\bR^3\times \bS^1$ as a function of the radius $R_4$ of $\bS^1$.  The
moduli space of the theory on $\bR^4$ has complex dimension one and has
an auxiliary torus fibered over it, which determines the metric on the
moduli space.  The moduli space of the theory on $\bR^3\times \bS^1$ is
the full four (real) dimensional fiber bundle where the area of the
auxiliary torus is $1 \over R_4$.

These facts can be derived by considering the effective theory of a
two-brane in compactification of M-theory on K3.  Since the brane is at
a point in K3, the moduli space of vacua of the theory on the brane is
K3.  For special values of the space time moduli (the parameters of the
theory) this moduli space becomes singular.  These singularities are
classified by an ADE classification.  Using this fact we can determine
the fate of the various singularities in four dimensions upon
compactification
\eqn\lowdnn{\eqalign{
A_n & \rightarrow A_n  \qquad {\rm for} ~~n \not= 0 \cr
D_n & \rightarrow D_n \qquad {\rm for} ~~n \ge 4 \cr
E_n &\rightarrow E_n\qquad  {\rm for} ~~n=6,7,8 \cr
H_n &\rightarrow A_n \qquad {\rm for} ~~n=1,2 .\cr}}
(The $A_0$ and $H_0$ points are nonsingular in dimension three.)

Notice that this is precisely the same mathematical phenomenon which was
responsible for the ambiguities in the lattice-theoretic specification
of F-theory models: only the singularity in the total space of the K3
surface, not the type of the singular fiber in the fibration, matters in
determining the gauge group (in the earlier example) or the
three-dimensional moduli space (in the present example).

In higher dimensions there are no known free field theories which flow
to the $E_n$ theories.  This fact has led some authors to suggest that
they are not local quantum field theories.  At least in three
dimensions, such free field theories were found in \intse.  Hence these
theories have a Lagrangian description and they are clearly local
quantum field theories.

One can generalize this discussion by considering the compactification
of the six-dimensional theory on $\bT^3$ as a function of its parameters
and background ($B_{\mu\nu}$ $\mu,\nu=4,5,6$).  All these parameters are
in vector multiplets and therefore, as explained in the introduction,
various non-renormalization theorems apply, e.g., the metric on the
Higgs branch is independent of these parameters.\foot{In three dimensions
vector multiplets are dual to hypermultiplets.  Nevertheless, the
non-renormalization theorems can be used because the scalars in these
multiplets transform under different global $SU(2)$ symmetries which are
exchanged by the duality transformation \intse.}  The appearance of
these theories as effective field theories on branes makes it obvious
that the resulting moduli space is again a piece of a K3 whose moduli
depend on the quark masses and the parameters of the compact $\bT^3$.
Therefore, by changing the parameters of the $\bT^3$ we can explore the
dynamics of these theories in various dimensions.  The answer is always
a piece of a K3.  It is amazing that the same object (K3) provides the
answer to so many different quantum field theories in different
dimensions!

\bigbreak\bigskip\bigskip
\centerline{\bf Acknowledgments}\nobreak
The work of O.J.G. was supported in part by a Robert H. Dicke fellowship
and by DOE grant DE-FG02-91ER40671; that of D.R.M. was supported in part by
the Harmon Duncombe Foundation and by NSF grants DMS-9401447 and
DMS-9627351; and that of N.S. was supported in part by DOE grant
DE-FG02-96ER40559.  We thank P. Berglund, B. Greene, A. Hanany, A. Losev,
G. Moore, V. Nikulin, Y. Oz, R. Plesser, S. Shatashvili, S. Shenker, and
E. Witten for helpful discussions.

%==================================================================%
%  Wilson lines                                                    %
%==================================================================%
\appendix{A}{Formulas for generic Wilson lines}
In section (3.2) we described the procedure to obtain the
4D Seiberg-Witten curve corresponding to compactification
with arbitrary $E_8$ twists along $\MT{2}$. The curve was
given in implicit form.
It is the purpose of this appendix to express the coefficients
of $f$ and $g$ as functions of the twists.

To describe the $E_8$ twists we decompose
\eqn\subs{
SO(2)^8 \subset SO(16) \longleftarrow
Spin(16) \subset E_8.}
and let  $\rho_1,\dots \rho_8$ be the values of the $SO(2)^8$ Wilson
lines around $R_6$ and let $\omega_1,\dots,\omega_8$ be the values
around $R_5$.

We define the complex variables:
\eqn\wils{
w_k = \omega_k + \rho_k\sigma,\qquad k=1\dots 8.}
These are 8 points on the torus.
The prepotential depends on them in a holomorphic manner,
subject to periodicity and to the $E_8$ Weyl group identifications:
\eqn\identif{\eqalign{
(w_1,\dots,w_8)
\sim &
(w_1+\half n_1 + \half m_1\sigma,\dots,w_8+\half n_8+ \half
m_8\sigma),\cr
&\,\,\,\,\,\,\qquad
n_i,m_i\in\IZ,\qquad \sum_1^8 n_i \equiv\sum_1^8 m_i \equiv 0 \pmod{2}
\cr
(w_1,\dots,w_8) \sim & (w_{\psi(1)}, \dots, w_{\psi(8)}),
\qquad \psi\in S_8
\cr
(w_1,\dots,w_8) \sim& ((-1)^{\epsilon_1}w_{1},
          \dots, (-1)^{\epsilon_8}w_{8}),
\qquad \sum_1^8 \epsilon_i \equiv 0 \pmod{2}
\cr
(w_1,\dots,w_8) \sim& (w_1 - {{\sum_1^8 w_i}\over 4},
 \dots, w_8 - {{\sum_1^8 w_i}\over 4}).
\cr
}}
With these Wilson lines the central charge formula is
\eqn\massa{
M = \sqrt{2}|n_e a + n_m a_D
 - 2\pi i v_5 (P_5 + \sum_{i=1}^8 S_i \omega_i)
 + 2\pi i v_6 (P_6 + \sum_{i=1}^8 S_i \rho_i)|.}
Here $P_5$ and $P_6$ are the momenta around $R_5$ and $R_6$
measured in integer units and $S_i$ are the the integer
$SO(2)$ charges in \subs.

We recall that $\omega_i$ and $\rho_i$, being Wilson lines,
were defined only up to integer shifts.
Indeed \massa\ is invariant under a change
\eqn\chga{\omega_i\rightarrow \omega_i +n_i,\qquad
          P_5\rightarrow P_5 -\sum_{i=1}^8 n_i S_i}
for integer $n_i$'s.

Also, under a modular transformation
\eqn\modulaa{R_5\leftrightarrow R_6, \,\,\varphi\rightarrow \pi
-\varphi,
\,\,\omega_i\rightarrow -\rho_i,\,\,\rho_i\rightarrow\omega_i,\,\,
 P_5\rightarrow -P_6, \,\,P_6\rightarrow P_5,\,\,}
the formula for the mass is still invariant provided we
change $a$ and $a_D$ according to
\eqn\modulaaad{a\rightarrow -i a_D,\,\,
a_D\rightarrow i a,\qquad n_e \rightarrow n_m,\,\,
n_m\rightarrow -n_e.}
which is consistent with \aadn.

%==================================================================%
%  More geometry                                                   %
%==================================================================%

The total space described by the elliptic fibration \ellip\
is the almost Del Pezzo surface with $\chi=12$ that also
appeared in the F-theory construction of $E_8$ tensionless
strings \MVII. An alternative, more convenient, description
of it is given by the blow-up of $\CP{2}$ at 9 points
$e_0,\dots, e_8$ which have to be the intersection of two
cubics. Let the homogeneous coordinates on $\CP{2}$ be $X,Y,Z$
and let the two cubics be
\eqn\twocub{P(X,Y,Z) = 0,\qquad Q(X,Y,Z) = 0.}
Then there is a whole family of cubics which intersect at
$e_0\dots e_9$:
\eqn\qup{Q(X,Y,Z) + u P(X,Y,Z) = 0}
where $u$ is a coordinate on $\CP{1}$.
The equation \qup\ exhibits the elliptic fibers of
$\CP{2}$ blown-up at the 9 points.

The cohomology structure is as follows \MVII:
the class of the fiber is
\eqn\fib{f= 3H -\sum_{i=1}^9 e_i}
where $H$ is the hyperplane section of $\CP{2}$.
The section of the elliptic fibration can be chosen to be the
exceptional divisor $e_0$ and the $E_8$ is generated by
\eqn\genee{
H-e_1-e_2-e_3,\, e_k-e_0.}
Now returning to \ellip\ we have to integrate the meromorphic
two-form $\Omega$ over the $E_8$ basis.
It will be more convenient to integrate the two-form over $e_i-e_0$.
The difference is just a linear combination (using the fact
that the integral of $\Omega$ on a fiber vanishes so that
we can subtract $\inv{3}f$ {}from $H-e_1-e_2-e_3$).
Now we need to identify $\Omega$ in the $\CP{2}$ variables.
This is done by noting that $\Omega$ has a single simple
pole on the whole fiber at $u=\infty$ which sets
$$
\Omega = {{d(X/Z)\,\wdg\, d(Y/Z)} \over {P(X/Z,Y/Z,1)}}.
$$
Now we have to calculate
$$
\int_{\lbr e_i \rbr - \lbr e_0 \rbr}\Omega.
$$
The $e_i$'s are analytic classes, so the integral of a general
$(2,0)$-form on them would vanish. However, the $e_i$ all lie
on the curve $P=0$ where $\Omega$ has a pole.
If we perform a linear change of variables to set
\eqn\pxyz{
P(X,Y,Z) = X^3 - \inv{4}g_2(\sigma) X Z^2 - \inv{4} g_3(\sigma) Z^3 - Y^2 Z}
where we used that fact that at $u=\infty$ the modulus of the cubic
has to be $\sigma$,
the integral turns out to be
$$
w_i = \int_{e_0}^{e_i} {dx' \over y'}
$$
where $x' = X/Z$ and $y' = Y/Z$.
In other words, under a map $\Phi$ {}from the elliptic curve $P=0$
to a lattice $\IC/\BZ + \BZ\sigma$ the points $e_i$
get mapped to $\Phi(e_0) + w_i$.
If we only knew what $\Phi(e_0)$ is we could have completed the
calculation. To determine that we note
$$
\sum_{i=0}^8 \Phi(e_i) = 0\pmod{\BZ + \BZ\sigma}.
$$
The standard proof is as follows \GriHar:
A general cubic in $\CP{2}$ intersects $P=0$ at 9 points so we can
define
a map {}from the space of cubics in $\CP{2}$ into $\IC/\BZ + \BZ\sigma$
by the sum of $\Phi$ of the intersection points.
The space of all cubics, however, is isomorphic to $\CP{9}$
and the only holomorphic map {}from $\CP{9}$ to a torus is a constant
because it is a constant when restricted to every line
$\CP{1}\subset\CP{9}$.

So we find that
\eqn\ew{\eqalign{
e_i =& (X={1\over \pi^2}\wp(\Phi(e_i)),
        Y={2\over \pi^3}\wp'(\Phi(e_i)), Z=1) \cr
w_i =& \Phi(e_i) - \Phi(e_0) \cr
0   =& \sum_{i=0}^8 \Phi(e_i) \cr
}}
Let's denote
\eqn\vi{\xi_i \equiv \Phi(e_i), \qquad i=0\dots 8.}

The $\xi_i$'s can easily be calculated, given the $w_i$'s.

So, given $\sigma$ we know $P(X,Y,Z)$ {}from \pxyz\ and given the $w_i$'s
we can calculate the $\xi_i$'s {}from \ew\ and then we can find
$Q(X,Y,Z)$ to be the cubic that passes through the $e_i$'s.

We write
\eqn\qxyz{Q(X_1, X_2, X_3) = \sum_{i,j,k}Q_{ijk} X_i X_j X_k}
where the $Q_{ijk}$ satisfy (putting $Q_{111}=0$ so as not to
get \pxyz\ back):
\eqn\eqsa{\eqalign{
0 =& Q_{333} + {8\over \pi^9} Q_{222} \wp'(\xi_i)^3
  + {6\over \pi^7} Q_{112}\wp(\xi_i)^2 \wp'(\xi_i)
  + {3\over \pi^4} Q_{113}\wp(\xi_i)^2 \cr
  &+ {12\over \pi^6} Q_{223} \wp'(\xi_i)^2
  + {12\over \pi^8} Q_{122} \wp'(\xi_i)^2 \wp(\xi_i)
  + {3\over \pi^2} Q_{133} \wp(\xi_i)\cr
  &+ {6\over \pi^3} Q_{233} \wp'(\xi_i)
  + {12\over \pi^5} Q_{123} \wp(\xi_i) \wp'(\xi_i)
\cr}}
Those are 8 equations $i=1\dots 8$ for 9 variables so there
is  one more solution.

The family of curves now has the form
\eqn\xyz{\eqalign{
0 =&  u (X^3 - f_4 X Z^2 - g_2 Z^3 - Y^2 Z)
    + (Q_{333} Z^3 + Q_{222} Y^3 + 3 Q_{112}X^2 Y + 3 Q_{113}X Y^2 \cr
  &+ 3 Q_{223} Y^2 Z + 3 Q_{122} Y^2 X + 3 Q_{133} X Z^2 + 3 Q_{233} Y
  Z^2
  + 6 Q_{123} X Y Z) . \cr}}
It can be changed into the form \ellip\ by an $SL(3,\IC)$
transformation.
The final step is to go back to the dimensionful variables
\eqn\goxy{
x = {1\over 8 v_5^2} \left({X\over Z}\right),\qquad
y = {1\over 16\sqrt{2} v_5^3} \left({Y\over Z}\right).
}

%------------------------------------------------------------------%
%  Decompactification                                              %
%------------------------------------------------------------------%

\appendix{B}{Decompactification from 4D to 5D}

The decompactification limit is reached by taking
$R_5\rightarrow\infty$ leaving $R_6$ fixed.
This means that $q\rightarrow 0$, but at the same time we must
scale $u$ according to
\eqn\scaleu{u = q^{-\psi}, \qquad
\psi = \sqrt{2}R_6\phi,
}
where $\psi$ is kept fixed.
The motivation for this is that in the large tension limit,
\eqn\ult{|u| \sim e^{4\pi^2 R_5 R_6 \times Tension}.}

In the decompactification limit we expect the 4D $U(1)$ coupling
constant to look like
\eqn\expu{
\tau = {8\pi i\over {g_{5}}^2} = i R_5 F(\psi)}
where $F(\psi)$ is a piecewise linear function of $\psi$ \fived.

Such a behavior will be obtained as follows.
For very large $\tau$ we can extract $\tau$ {}from \ellip:
$$
{{f^3}\over {g^2}} =  {{27} \over {4}} + 11664 e^{2\pi i\tau} + \cdots
$$
On the other hand, $f$ and $g$ are given by
\eqn\fgx{
f = \sum_{j=0}^4 f_j u^j,\qquad g = \sum_{l=0}^6 g_l u^l.
}
 For small $q$ the coefficients $f_4$ and $g_6$
are:
\eqn\fgsq{\eqalign{
f_4 =&
\inv{256 {v_5}^4}  g_2(\sigma)
=\inv{256 {v_5}^4}
\left({1\over 3} + 80 q + \ol{q^2}\right)
\cr
g_6 =&
\inv{2048 {v_5}^6} g_3(\sigma)
=\inv{2048 {v_5}^6}
\left({2\over 27} - {112\over 3} q + \ol{q^2}\right)
\cr
}}
The other coefficients will also behave like $q^{-\a_k}$
for some  powers $\a_k$ which depend on the Wilson lines.
Thus we will find an expression of the form
\eqn\domin{
{{f^3}\over {g^2}} -  {{27} \over {4}} =
          11664 q + c_1 q^{\psi -\a_1} + c_2 q^{2\psi -\a_2} + \cdots}
For very large $\psi$ the first term on the right is dominant
as $q\rightarrow 0$ giving a constant $\tau$ in 5D.
As we decrease $\psi$ we will reach a value for which one of the other
terms becomes dominant. This way we see that $\tau$ is a piecewise
linear function of $\psi$ with jumps in the derivative where
we switch {}from one term to the other in \domin.

%------------------------------------------------------------------%
%  Unbroken E8                                                     %
%------------------------------------------------------------------%
\subsec{Decompactification for unbroken $E_8$}

Now let's take the limit $q\rightarrow 0$ in \unbrk.
Substituting \scaleu\ in \unbrk\ and using \fgsq\ we find:
\eqn\coupl{
\tau = \cases{
\sigma          & for~ $\psi > 1$ \cr
\psi\sigma+ \ol{\log |\sigma|}
                     & for~ $1 > \psi > 0$ \cr
\half + i{\sqrt{3}\over 2}    & for~ $0 > \psi$ \cr}}
The 5D coupling constant is thus:
\eqn\gfivd{
{8\pi \over {g_5}^2} = \,\,\lim_{R_5\rightarrow\infty}
   \left({\tau\over 2\pi i R_5}\right)
=\cases{
{1\over 2\pi R_6}       & for~ $\psi > 1$ \cr
{1 \over 2\pi R_6}\psi  & for~ $1 > \psi > 0$ \cr
0                       & for~ $0 > \psi$ ~. \cr}}
which agrees with \gefffs.

Let us see where the singularities are.
Solving for the roots of the discriminant we find
10 singularities at $u=0$ which is formally $\psi=-\infty$
and two additional singularities at
\eqn\soldit{
u\sim -{27i\over 2} \sigma^3,\qquad
  {i\over 64}\sigma^3 q^{-1}
}
The first one is pushed to the boundary $\psi=0$ in 5D while
the second one becomes the single singularity at $\psi=1$.

%------------------------------------------------------------------%
%  The general case                                                %
%------------------------------------------------------------------%
\subsec{The general case}

For the general case, with Wilson lines, we have to use the curve \xyz.

The $\tau$ of the corresponding torus can be calculated {}from
the formula for the $j$-invariant of a curve in the form
\eqn\ccc{\sum_{1\le i,j,k\le 3} C_{ijk} X_i X_j X_k = 0.}
The result is an $SL(3,\IC)$ invariant rational function
of the $C_{ijk}$'s  and is given diagrammatically by:

\iffigs
\epsfxsize=80mm
\centerline{ \epsfbox{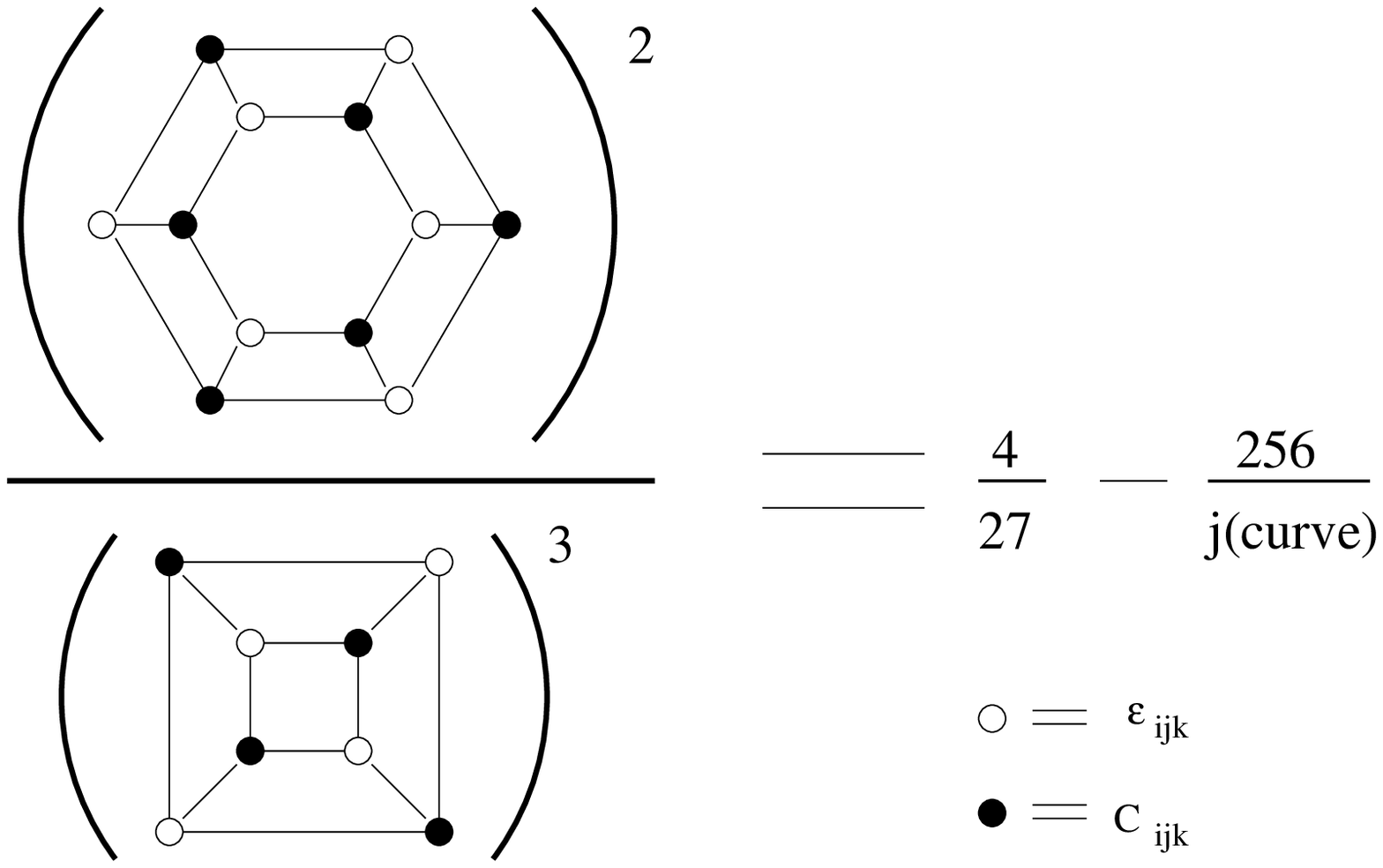} }\nobreak
\centerline{Fig.A: The $j$-function of the general cubic.}
\else \vskip1in
\fi

\noindent
where a hollow circle is $\epsilon_{ijk}$ and a full circle is
$C_{ijk}$ and lines denote index contractions.

In \xyz\ we use the $\xi_i$'s which are linear combinations of the
$w_i$'s according to \ew.
The 5D Wilson lines are along the $R_6$ direction so we have
\eqn\wkf{w_k = \inv{2\pi i}\rho_k \log q}
and
\eqn\gk{\xi_k = \inv{2\pi i}\g_k \log q}
where the $\g_k$ are linear combinations of the real $\rho_k$
according to \ew.
For small $q$ and assuming $-\half <\g_k < \half$ we expand:
\eqn\pwsmallr{\eqalign{
\inv{(2\pi i)^2}\wp(\xi_k) =& \inv{12} +{q^{\g_k}\over {(1-q^{\g_k})^2}}
+\ol{q^{1-\g_k}} \cr
\inv{(2\pi i)^3}\wp'(\xi_k) =& {q^{\g_k}(1+q^{\g_k})\over
{(1-q^{\g_k})^3}} +\ol{q^{1-\g_k}} .\cr
}}

The set of equations \eqsa\ becomes:
\eqn\neweqs{\sum_{n=0}^9 B_n q^{n \g_k} = 0,\qquad k=1\dots 8}
where the $B_k$-s are appropriate  linear combinations of the
$Q_{ijk}$-s.

We will assume:
$$
 |\g_1| > |\g_2| > \dots > |\g_8| .
$$
We find the approximate solution:
\eqn\solbs{B_0 = -1,\qquad |B_k| = q^{-|\g_8|-|\g_7| -\cdots -|\g_{8-k}|
  -k|\g_{9-k}|},\qquad k=1\dots 8.}

Substituting \solbs\ into \xyz\ and (Fig.A) we find
that the dominant terms are
\eqn\taudom{
e^{2\pi i\tau} =
q + \sum_{j=1}^9 c_j B_{j-1} B_8^{j-1} u^{-j}
= \sum_{j=0}^{9} c_j q^{k\psi - \a_k}.
}
with
\eqn\alphas{\eqalign{
\a_0 &= -1,\qquad \a_1 = 0,\cr
\a_k &= 8(k-1)|\gamma_1| + (k-1)\sum_{2\le j\le 9-k}|\gamma_j| +
         2(k-1)|\gamma_{10-k}| + k \sum_{11-k\le j\le 8} |\gamma_j| ,
       \qquad k=2\dots 8\cr
\a_9 =&
9|\g_8| +9|\g_7| +9|\g_6| +9|\g_5| +9|\g_4| +9|\g_3| +9|\g_2|
+72|\g_1|.\cr
}}
Now we can read of the coupling constant {}from \taudom.
It is (see Fig.B)
\eqn\ggen{
{8\pi i\over {g_5}^2} = \left({1 \over 2\pi R_6}\right)\,\,
            \min_{k=0\dots 9} (k \psi  - \a_k),
\qquad \psi = \sqrt{2} R_6\phi.
}
This is a piecewise linear function.

\iffigs
\epsfxsize=80mm
\centerline{ \epsfbox{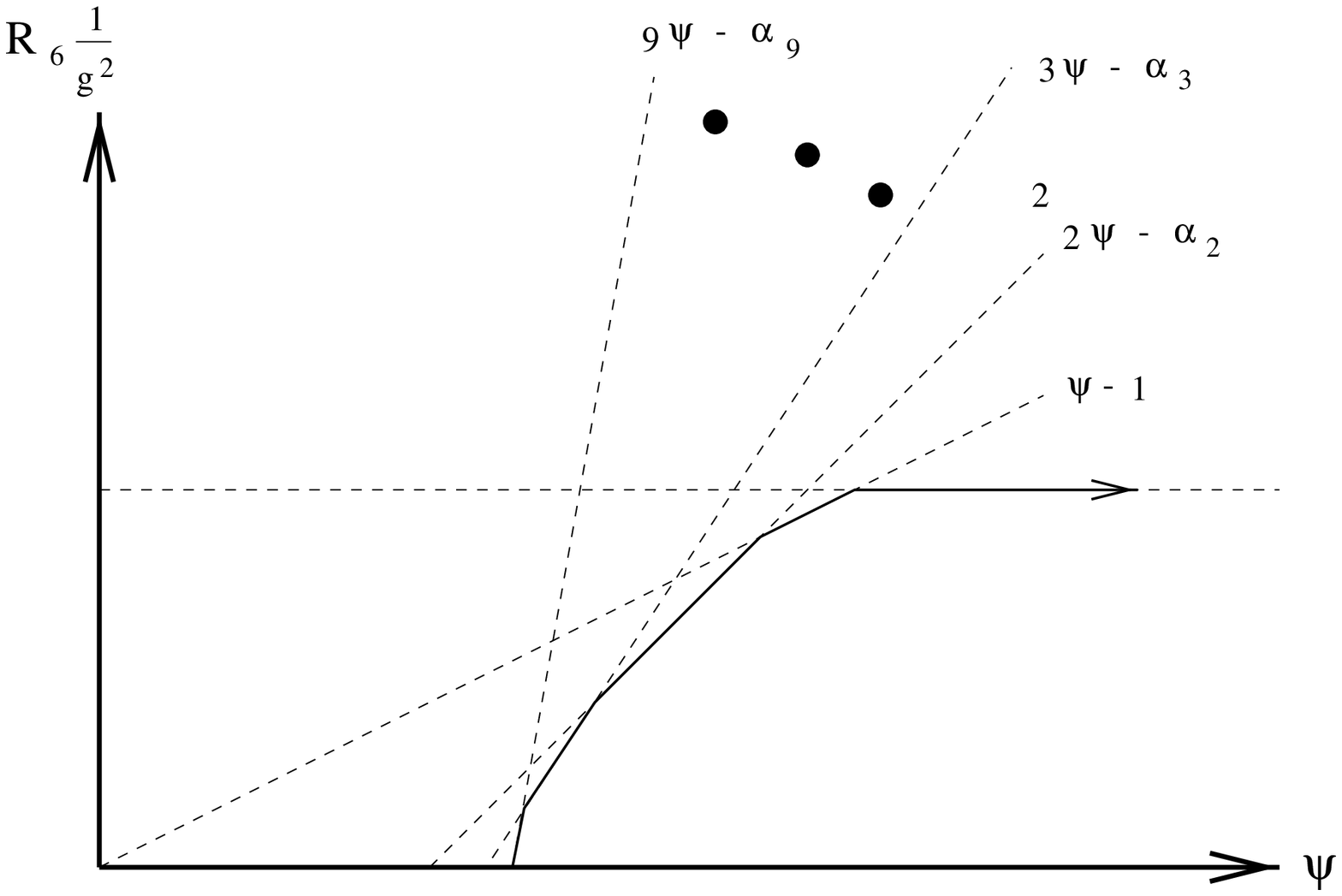} }\nobreak
\centerline{Fig.B: The 5D coupling constant
                    is a piecewise linear function of $\psi$.}
\fi

The number of discontinuities of the derivative could be anything
between 1 and 9 according to the precise values of the $\a_k$'s.

Formula \ggen\ describes the general case for small Wilson lines.
It includes the cases discussed in the previous subsections
but it is not the {\it most} general case.
When the $\g_k$'s are not necessarily all small, we should
take account of more terms in the expansion of the $\wp$-functions
in \pwsmallr.

\listrefs
\end